\documentclass[aps,prb,twocolumn,reprint,superscriptaddress,citeautoscript]{revtex4-2}
\usepackage[dvipdfmx]{graphicx}
\usepackage{subfigmat}
\usepackage{amsmath,amssymb}
\usepackage{color}
\usepackage{bm}
\usepackage{hyperref}
\usepackage{comment}
\usepackage[normalem]{ulem}
\usepackage{cancel}
\usepackage{upgreek}
\usepackage{braket}

\begin{document}

\title{Microscopic theory of Rashba-Edelstein magnetoresistance}

\author{Masaki Yama}
\affiliation{Institute for Solid State Physics, The University of Tokyo, 5-1-5 Kashiwanoha, Kashiwa, 277-8581 Japan}

\author{Mamoru Matsuo}
\affiliation{Kavli Institute for Theoretical Sciences, University of Chinese Academy of Sciences, Beijing, China}
\affiliation{CAS Center for Excellence in Topological Quantum Computation, University of Chinese Academy of Sciences, Beijing, China}
\affiliation{Advanced Science Research Center, Japan Atomic Energy Agency, Tokai, Japan}
\affiliation{RIKEN Center for Emergent Matter Science (CEMS), Wako, Saitama, Japan}

\author{Takeo Kato}
\affiliation{Institute for Solid State Physics, The University of Tokyo, 5-1-5 Kashiwanoha, Kashiwa, 277-8581 Japan}

\date{\today}

\begin{abstract}
We theoretically study Rashba-Edelstein magnetoresistance (REMR) in a two-dimensional electron gas (2DEG) system with Rashba and Dresselhaus spin-orbit interactions. We consider a microscopic model of a junction system composed of a ferromagnetic insulator and a 2DEG, and derive analytic expressions for the spin and current densities in the 2DEG using the Boltzmann equation, while taking into account dynamical contributions from magnons. Our findings reveal that the sign of the REMR varies depending on the type of interface. We also discuss the experimental relevance of our results.
\end{abstract}
\maketitle 

\section{Introduction}
\label{sec:introduction}

The phenomena of spin-to-charge and charge-to-spin conversion have been investigated extensively in the development of spintronic devices. 
Spin Hall effect (SHE)~\cite{Kato2004, Wunderlich2005, Sinova2015} and its reciprocal effect, the inverse spin Hall effect (ISHE)~\cite{Saitoh2006}, are such prototypical examples.
These two conversion phenomena cause spin Hall magnetoresistance (SMR)~\cite{Nakayama2013, Kato2020, Ishikawa2023}, which manifests in ferromagnetic material/heavy metal junctions. Meanwhile, at interfaces in junction systems, the structure inversion asymmetry gives rise to the Rashba spin-orbit interaction~\cite{Bychkov1984, Rashba2015,Winkler2003}, which is an increasingly recognized effect influencing spin-charge conversion and magnetoresistance phenomena~\cite{Kobs2011, Hupfauer2015, Olejnik2015, Kobs2016, Ado2017, Kang2020, Kundu2022, Comstock2023, Chen2023rev}.

The Rashba spin-orbit interaction, along with the Dresselhaus spin-orbit interaction inherent in the Zinc-Blende crystal structure~\cite{Dresselhaus1955, Winkler2003}, causes spin-splitting in the energy bands of the two-dimensional electron gas (2DEG) formed near the interface, leading to spin-momentum locking~\cite{Manchon2015, Kohda2017}. 
In such 2DEG, the charge-to-spin conversion phenomenon known as the Rashba-Edelstein effect (REE)~\cite{Edelstein1990, Inoue2003, Silsbee2004, Trushin2007, Suzuki2023, Johansson2024} or its inverse effect referred to as the inverse Rashba-Edelstein effect (IREE)~\cite{Ganichev2002, Sanchez2013, Shen2014a, Suzuki2023, Yama2023b, Hosokawa2024} has been observed. 
The magnetoresistance effect that arises from the interplay of these two conversion phenomena at the interface is termed Rashba-Edelstein magnetoresistance (REMR)~\cite{Nakayama2016,Nakayama2017,Nakayama2018,Thompson2020}, while different terms have been used for the same phenomenon in other references~\cite{JKim2017,Du2021,Narayanapillai2017,Zhou2018,Grigoryan2014,footnote}.
REMR has been observed in various junction systems, such as Bi/Ag/CoFeB~\cite{Nakayama2016, Nakayama2017, Nakayama2018}, CoFe/Cu/${\rm Bi_{2}O_{3}}$~\cite{JKim2017}, Pt/Co~\cite{Thompson2020, Du2021}, LAO/STO~\cite{Narayanapillai2017}, Cu[Pt]/YIG~\cite{Zhou2018}, YIG/atomic layer materials~\cite{Mendes2021}, and Cr/YIG~\cite{Chen2023}. Furthermore, several theoretical studies have been conducted on this topic~\cite{Grigoryan2014, Zhang2014, Zhang2015b, Iguchi2017, Tolle2018a, Tolle2018b, Sanz-Fernandez2020, Silva2023}.
However, previous theoretical studies treat spin transfer at the interface phenomenologically using spin-mixing conductance, limiting their predictive capability regarding the REMR such as detailed dependence of charge current modulation on the magnetization orientation of the ferromagnet for a complex spin-splitting Fermi surface.

\begin{figure}[tb]
\begin{center}
\includegraphics[width=85mm]{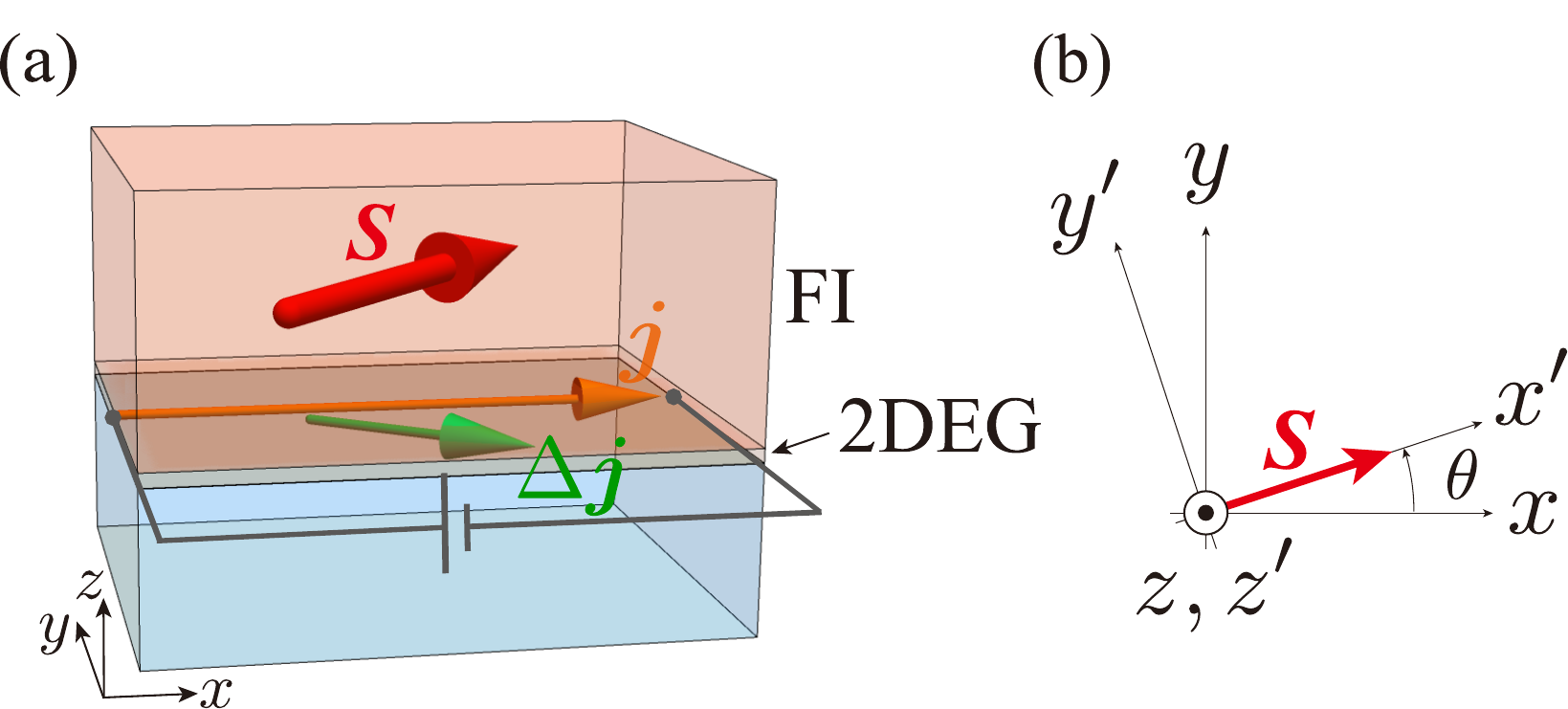}
\caption{(a) A schematic picture of an experimental setup considered in this study. 
The red arrow represents a spontaneous spin polarization of the FI, $\bm{S}$, while the orange arrow represents a charge current
$\bm{j}$ induced by an external electric field.
The current modulation $\Delta\bm{j}$ (indicated by the green arrow) is caused by a combination of the direct and inverse Rashba-Edelstein effects.
(b) Relation between the laboratory coordinates $(x,y,z)$ and the $\bm{S}$-fixed coordinates $(x',y',z')$.}
\label{fig:setup}
\end{center}
\end{figure}

In this study, we theoretically formulate REMR for a junction system composed of a ferromagnetic insulator (FI) and 2DEG with Rashba and Dresselhaus spin-orbit interactions (see Fig.~\ref{fig:setup}).
We express spin transfer at the interface through a microscopic Hamiltonian~\cite{Ohnuma2014, Matsuo2018, Kato2019, Kato2020, Ominato2020a, Ominato2020b, Ominato2022, Funato2022, Tajima2022, Yama2021, Yama2023, Yama2023b, Hosokawa2024}, which allows us to analyze the detailed behavior of REMR without using phenomenological parameters. Assuming that the spin-orbit interaction energy is much larger than the impurity scattering rate, we calculate non-equilibrium distribution functions of conduction electrons in 2DEG under an external electric field by the Boltzmann equation~\cite{Suzuki2023, Yama2023b, Hosokawa2024, Kato2024}. We should note that a part of our calculation is analogous as previous microscopic calculations using the Boltzmann equation for the anisotropic magnetoresistance (AMR) in diluted magnetic semiconductors~\cite{Vyborny2009,Trushin2009},  where an effective transverse Zeeman field on conduction electrons induced by magnetic impurities was taken into account.
Although this transverse field effect is analogous to that induced by an exchange bias at an interface in our calculation, their theory cannot describe dynamical effect due to magnon absorption and emission, highlighting the need for a different theoretical framework.

A key feature of our study is to construct a microscopic theory, which enables us to predict the dependence of REMR on the magnetization direction and to clarify the effects of magnons in the FI and interfacial randomness on REMR.
We describe REMR by introducing the interfacial collision term accompanying magnon annihilation or creation and consider two types of the FI-2DEG interface, i.e., clean and dirty interfaces.
We show that the sign of REMR determined from the spin-orientation dependence in the FI is different for these two types of interfaces.
We also discuss how the amplitude of the REMR depends on the ratio between the strength of the Rashba and Dresselhaus spin-orbit interactions.
Our result for the dirty interface is consistent with the experiment for the Bi/Ag/CoFeB junction system~\cite{Nakayama2016}. 
We present the physical mechanism behind the positive and negative REMR and discuss its relevance to experiments.

The rest of this work is organized as follows. The model Hamiltonians of the 2DEG/FI bilayer system are first presented in Sec.~\ref{sec:model}.
REMR is formulated in Sec.~\ref{sec:formulation} and is calculated for several parameters in Sec.~\ref{sec:results}. 
The experimental relevance of our results is discussed in Sec.~\ref{Sec:Experiment}.
Our results are summarized in Sec.~\ref{sec:summary}.
The appendices are devoted to explicit expressions of the collision terms, detailed calculations of spin and charge densities, and the results of nonequilibrium distribution functions.

\section{Model}
\label{sec:model}

In this section, we introduce a model for the 2DEG-FI junction system as shown in Fig.~\ref{fig:setup}.
The Hamiltonians for 2DEG, FI, and the interface between them are given in Sec.~\ref{sec:2DEG}, \ref{sec:FI}, and \ref{sec:interface}, respectively.

\subsection{Two-dimensional electron gas (2DEG)}
\label{sec:2DEG}

The model Hamiltonian of 2DEG is given as follows:
\begin{align}
& H_{\rm kin}=\sum_{\bm{k}}
\begin{pmatrix}
c^{\dagger}_{\bm{k}\uparrow} & c^{\dagger}_{\bm{k}\downarrow} 
\end{pmatrix}
\hat{h}_{\bm{k}}
\begin{pmatrix}
c_{\bm{k}\uparrow} \\ c_{\bm{k}\downarrow} 
\end{pmatrix},\label{eq:defHkin} \\
& \hat{h}_{\bm{k}}=
(\epsilon_{\bm k} -\mu)\hat{I}+\alpha(k_y\hat{\sigma}_x-k_x\hat{\sigma}_y)+\beta(k_x\hat{\sigma}_x-k_y\hat{\sigma}_y),
\end{align}
where $c_{{\bm k}\sigma}^\dagger$ ($c_{{\bm k}\sigma}$) are the creation (annihilation) operators for conduction electrons with wavevector $\bm{k}=(k_x,k_y)^T$ and spin $\sigma$ ($=\uparrow, \downarrow$).
The energy dispersion is given as $\epsilon_{\bm{k}} = \hbar^2(k_x^2+k_y^2)/2m^*$ ($m^*$: an effective mass), $\mu$ is a chemical potential, and the magnitudes of the Rashba and Dresselhaus spin-orbit interactions are denoted by $\alpha$ and $\beta$, respectively. 
We denote the $2\times 2$ identity matrix and the Pauli matrices by
$\hat{I}$ and $\hat{\bm \sigma}=(\hat{\sigma}_x, \hat{\sigma}_y)^{T}$, respectively. 
Using the polar representation as ${\bm k}=(|{\bm k}|\cos\varphi,|{\bm k}|\sin \varphi)^T$, the $2\times 2$ matrix $\hat{h}_{\bm{k}}$ is rewritten as
\begin{align}
&\hat{h}_{\bm{k}}= (\epsilon_{\bm k}-\mu)\hat{I}-\bm{h}_{\rm eff}\cdot\hat{\bm{\sigma}},\\
&\bm{h}_{\rm eff}(\bm{k})
=|\bm{k}|\left(\begin{array}{c}
-\alpha\sin\varphi-\beta\cos\varphi \\
\alpha\cos\varphi+\beta\sin\varphi \end{array} \right).
\label{eq:Zeeman}
\end{align}
Here, $\bm{h}_{\rm eff}=(h_x,h_y)^T$ is an effective Zeeman field acting on conduction electrons.

Hereafter, we assume that the spin-orbit interaction energies, $k_{\rm F}\alpha$ and $k_{\rm F}\beta$, are much smaller than the Fermi energy.
Then, the effective Zeeman field can be approximated as
\begin{align}
\bm{h}_{\rm eff}(\bm{k})
&\simeq k_{\rm F}\left(\begin{array}{c}-\alpha\sin\varphi-\beta\cos\varphi \\ \alpha\cos\varphi+\beta\sin\varphi \end{array} \right), 
\label{eq:Zeeman2}
\end{align}
where $k_{\rm F}$ is the Fermi wavenumber in the absence of the spin-orbit interactions defined as $\epsilon_{\rm F}=\hbar^2 k_{\rm F}^2/2m^*$, where $\epsilon_{\rm F}$ is the Fermi energy, i.e., the zero-temperature chemical potential.
Because $\bm{h}_{\rm eff}(\bm{k})$ depends only on the orientation of the wavevector of conduction electrons, $\varphi$, we denote the effective Zeeman field as ${\bm h}_{\rm eff}(\varphi)$ hereafter.

In the presence of the Rashba and Dresselhaus spin-orbit interactions, the electronic energy bands in the 2DEG are split into two spin-polarized subbands, whose energy dispersion is given by
\begin{align}
E^{\gamma}_{\bm{k}}&=\epsilon_{\bm{k}}+\gamma h_{\rm eff}(\varphi), \label{eq:Egk}
\end{align}
where 
\begin{align}
h_{\rm eff}(\varphi)&\equiv |\bm{h}_{\rm eff}(\varphi)|= k_{\rm F}\kappa(\varphi),\\
\kappa(\varphi)&\equiv \sqrt{\alpha^2+\beta^2+2\alpha\beta\sin2\varphi}, \label{def:kappa}
\end{align}
and $\gamma$ ($=\pm$) labels the two subbands. The corresponding eigenstates are expressed as
\begin{align}
|\bm{k}\gamma\rangle 
&= \frac{1}{\sqrt{2}}
\begin{pmatrix}
C(\varphi) \\ \gamma
\end{pmatrix},\label{eq:eigenkg}\\
C(\varphi)&\equiv \frac{-h_x(\varphi)+ih_y(\varphi)}{h_{\rm eff}(\varphi)}.
\end{align}
Using these eigenstates, the relationship between the annihilation operators for the $\sigma$ ($=\uparrow,\downarrow$) basis and the $\gamma$ ($=\pm$) basis in the 2DEG can be established as follows:
\begin{align}
c_{\bm{k}\sigma}& =\sum_{\gamma}C_{\sigma\gamma}(\varphi)c_{\bm{k}\gamma}, \label{eq:ckstg}
\end{align}
where $C_{\uparrow \gamma}= C(\varphi)/\sqrt{2}$, and $C_{\downarrow \gamma} = \gamma/\sqrt{2}$. 

We also consider impurity scattering by the Hamiltonian
\begin{align}
H_{\rm imp}&=\sum_i \sum_{\sigma}
\int d\bm{r} \, v(\bm{r}-\bm{R}_i)\psi^{\dagger}_{\sigma}(\bm{r})\psi_{\sigma}(\bm{r})
\end{align}
where  
$\psi_{\sigma}(\bm{r})=\mathcal{A}^{-1/2} \sum_{\bm{k}}e^{i\bm{k}\cdot\bm{r}}c_{\bm{k}\sigma}$, $v(\bm{r})$ is an impurity potential, $\bm{R}_i$ denotes an impurity position, and $\mathcal{A}$ is an area of the interface.
For simplicity, we consider point-like impurities modeled by $v(\bm{r})=u\delta({\bm r})$, where $u$ denotes a strength of the impurity potential and $\delta({\bm r})$ is the delta function.
The magnitude of impurity scattering is quantified by energy broadening $\Gamma = 2 \pi n_{\rm imp} u^2 D(\epsilon_{\rm F})$, where $n_{\rm imp}$ is the impurity concentration and $D(\epsilon_{\rm F})$ is the density of states at the Fermi energy per unit area per spin.
Throughout this study, we assume the condition $\Gamma \ll {\rm max}(k_{\rm F}\alpha, k_{\rm F}\beta )$, for which spin-momentum locking in 2DEG is most effective.

\subsection{Ferromagnetic insulator (FI)}
\label{sec:FI}

We describe the FI by the Heisenberg model
\begin{align}
H_{\rm FI} &= \sum_{\langle i,j \rangle}
J_{ij} {\bm S}_i \cdot {\bm S}_j -\hbar \gamma_{\rm g} \sum_{i} {\bm h}_{\rm dc}\cdot {\bm S}_{i}, \label{eq:HFI1} \\
{\bm h}_{\rm dc} &= (-h_{\rm dc} \cos \theta, -h_{\rm dc} \sin \theta, 0), 
\end{align}
where $J_{ij} (<0)$ represents the ferromagnetic exchange interaction, $\langle i,j \rangle$ denotes nearest neighbor pairs, $\gamma_{\rm g} (<0)$ the gyromagnetic ratio, and $\bm{h}_{\rm dc}$ is an external static magnetic field with $\theta$ being the angle of this field. 
For simplicity, we neglect the magnetic anisotropy~\footnote{It is not difficult to include anisotropy terms, such as uniaxial magnetic anisotropy, in our formulation. Such terms would primarily modify the magnon dispersion relation and could quantitatively influence the temperature dependence of the results, in particular, at low temperatures. Since a detailed analysis of the temperature dependence is not the primary focus of this study, we neglect these anisotropy terms in our study.}.
We assume that the temperature is much lower than the magnetic transition temperature.
The expectation value of the spin polarization in FI is expressed as $\langle \bm{S}_i\rangle = (\langle S^x_i\rangle, \langle S^y_i\rangle, \langle S^z_i\rangle)=(S_0\cos\theta, S_0\sin\theta, 0)$. 
We further employ the spin-wave approximation, assuming that the magnitude of spin $S_0$ is much larger than unity.
In applying the spin-wave approximation, it is convenient to introduce a new coordinate $(x',y',z')$, in which the $x'$ axis is fixed with the direction of the spin polarization of the FI.
We note that in this new coordinate, the expectation value of the spin is given as $\langle \bm{S}_i\rangle = (\langle S^{x'}_i\rangle, \langle S^{y'}_i\rangle, \langle S^{z'}_i\rangle)=(S_0,0,0)$ (see Fig.~\ref{fig:setup}(b)). 
The spin operators expressed in these two coordinates are related to each other as
\begin{align}
\begin{pmatrix}
S^{x'}_i \\ S^{y'}_i \\ S^{z'}_i
\end{pmatrix}
= \begin{pmatrix}
\cos\theta & \sin\theta & 0
\\ -\sin\theta & \cos\theta & 0 
\\ 0 & 0 & 1
\end{pmatrix}
\begin{pmatrix}
S^{x}_i \\ S^{y}_i \\ S^{z}_i
\end{pmatrix}.
\end{align}
Utilizing the Holstein-Primakoff transformation
\begin{align}
S^{x'+}_j &= S^{y'}_j + iS^{z'}_j \simeq (2S_0)^{1/2} b_j,\label{eq:HolS+} \\
S^{x'-}_j &= S^{y'}_j - iS^{z'}_j \simeq (2S_0)^{1/2} b_j^\dagger,\label{eq:HolS-}\\
S^{x'}_j &= S_0 - b_j^\dagger b_j,\label{eq:HolSxp}
\end{align}
and the Fourier transformation of the magnon annihilation operator
$b_j = N_{\rm FI}^{-1/2} \sum_{\bm {q}} e^{i{\bm q}\cdot {\bm r}_j} b_{\bm q}$, the Hamiltonian of the FI is given in the leading order of $1/S_0$ as
\begin{align}
H_{\rm FI} &= \sum_{\bm{q}}\hbar\omega_{\bm{q}} b^{\dagger}_{\bm{q}}b_{{\bm q}},\\
\hbar\omega_{\bm{q}} &= \mathcal{D}\bm{q}^2+\hbar|\gamma_{\rm g}| h_{\rm dc},
\end{align}
where ${\bm q}=(q_x,q_y,q_z)$ represents the three-dimensional wavenumber of magnons, $N_{\rm FI}$ denotes the number of unit cells in the FI, $\omega_{\bm{q}}$ represents the dispersion relation, and $\mathcal{D}$ is a spin stiffness.

\subsection{FI/2DEG interface}
\label{sec:interface}

In the laboratory frame, the spin operators for conduction electrons in 2DEG are expressed as
\begin{align}
s^a_{\bar{\bm{q}}}&= \sum_{\sigma,\sigma'}\sum_{\bm{k}}c^{\dagger}_{\bm{k}\sigma}(\hat{\sigma}_a)_{\sigma\sigma'}c_{\bm{k}+\bar{\bm{q}}\sigma}, \quad (a= x, y, z),
\end{align}
where $\bar{\bm q}=(\bar{q}_x,\bar{q}_y)$ is a two-dimensional wavenumber, $\hat{\sigma}_a$ denotes the Pauli matrices for $a=x, y, z$. The spin operators in this frame are related to those in the transformed coordinate system $(x', y', z')$ as follows:
\begin{align}
\begin{pmatrix}
s^{x'}_i \\ s^{y'}_i \\ s^{z'}_i
\end{pmatrix}
&=
\begin{pmatrix}
\cos\theta & \sin\theta & 0
\\ -\sin\theta & \cos\theta & 0 
\\ 0 & 0 & 1
\end{pmatrix}
\begin{pmatrix}
s^{x}_i \\ s^{y}_i \\ s^{z}_i
\end{pmatrix} .
\end{align}
By the Fourier transformation, the spin ladder operators in the new coordinate system $(x', y', z')$ 
are expressed as
\begin{align}
s^{x'}_{\bar{\bm{q}}}&=\frac{1}{2}\sum_{\sigma,\sigma'}\sum_{\bm{k}}c^{\dagger}_{\bm{k}\sigma}(\hat{\sigma}^{x'})_{\sigma\sigma'}c_{\bm{k}+\bar{\bm{q}}\sigma'},\label{def:sxpq} \\
s^{x'\pm}_{\bar{\bm{q}}}
&=\frac{1}{2}\sum_{\sigma,\sigma'}\sum_{\bm{k}}c^{\dagger}_{\bm{k}\sigma}(\hat{\sigma}^{x'\pm})_{\sigma\sigma'}c_{\bm{k}\pm\bar{\bm{q}}\sigma'},\label{def:sxppmq}
\end{align}
where $\hat{\sigma}^{x'}$ and $\hat{\sigma}^{x'\pm}$ are written with the Pauli matrices $\sigma_a$ ($a=x,y,z$) as
\begin{align}
\hat{\sigma}^{x'}&=\cos\theta~\hat{\sigma}_{x}+\sin\theta~\hat{\sigma}_{y},\\
\hat{\sigma}^{x'\pm}
&=-\sin\theta~\hat{\sigma}_x +\cos\theta~\hat{\sigma}_y \pm i\hat{\sigma}_z.\label{eq:hatsig}
\end{align}
Using these spin operators, the interfacial exchange coupling at a FI/2DEG interface is generally described by the following Hamiltonian~\cite{Ohnuma2014,Matsuo2018, Kato2019,Kato2020,Ominato2020a, Ominato2020b,Ominato2022,Funato2022, Tajima2022,Yama2021,Yama2023,Yama2023b,Hosokawa2024}:
\begin{align}
H_{\rm int} &=H_{\rm int,d}+H_{\rm  int,s},\label{eq:Hd+Hs}\\
H_{\rm int,d}&=\sum_{{\bm q}} \sum_{\bar{\bm{q}}}(T_{\bm{q},\bar{\bm{q}}} S^{x'+}_{\bm{q}} s^{x'-}_{\bar{\bm{q}}}
+T^{*}_{\bm{q},\bar{\bm{q}}} S^{x'-}_{\bm{q}} s^{x'+}_{\bar{\bm{q}}}), \\
H_{\rm int,s}&=\sum_{\bar{\bm{q}}}\mathcal{T}_{\bm{0},\bar{\bm{q}}}
S_0 s^{x'}_{\bar{\bm{q}}},\label{eq:Hintdirty}
\end{align}
where $T_{\bm{q},\bar{\bm{q}}}$ and $\mathcal{T}_{\bm{0},\bar{\bm{q}}}$ represent the strengths of the exchange interactions.
$H_{\rm int,d}$ addresses magnon absorption and emission processes of the interface, while $H_{\rm int,s}$ describes the effect of the exchange bias, that is, the effective Zeeman field felt by conduction electrons in 2DEG.

In our study, we consider two types of the FI/2DEG interface, i.e., a dirty and clean interface:
\begin{align}
{\rm dirty \ interface:}& \ T_{{\bm q},\bar{\bm q}}= \bar{T}_{q_z}, \ {\cal T}_{{\bm 0},\bar{\bm q}}=\bar{\cal T}_{q_z}, \label{eq:dirty} \\
{\rm clean \ interface:}& \ T_{{\bm q},\bar{\bm q}}=\bar{T}_{q_z}
\delta_{{\bm q}_{\parallel},\bar{\bm q}}, \ {\cal T}_{{\bm 0},\bar{\bm q}}=\bar{\cal T}_{q_z} \delta_{\bar{\bm q},{\bm 0}}, \label{eq:clean}
\end{align}
where ${\bm q}_{\parallel}=(q_x,q_y)$ is an in-plane component of the magnon wavenumber ${\bm q}$, $q_z$ is $z$ component of $\bm{q}$, and $\bar{T}_{q_z}$ and $\bar{\mathcal{T}}_{q_z}$ are the interfacial exchange couplings. We note that the clean interface conserves the in-plane momentum while the dirty one does not. The explicit forms of $\bar{T}_{q_z}$ and $\bar{\mathcal{T}}_{q_z}$ depend on a detail of the modeling of the interface. If we consider a clean FI with a finite thickness and assume spatially uniform interfacial couplings, they are determined as
\begin{align}
\bar{T}_{q_z} &= \bar{T} \sin (q_z a), \label{tqz0} \\
\bar{\mathcal{T}}_{q_z} &= \bar{\mathcal{T}},
\label{tqz}
\end{align}
where $\bar{T}$ and $\bar{\mathcal{T}}$ are constants (for a detailed derivation, see Appendix~\ref{app:DrvHint}).
Although we employ this simple model in the subsequent calculations, the forms of $\bar{T}_{q_z}$ and $\bar{\mathcal{T}}_{q_z}$ do not affect the qualitative features of the results shown later; they modify only the amplitude and the temperature dependence of the REMR, which will not be discussed in details.

\section{Formulation}
\label{sec:formulation}

In this section, we formulate the Rashba-Edelstein magnetoresistance (REMR) in an FI/2DEG junction system.
We first introduce the Boltzmann equation in Sec.~\ref{sec:BoltzmannEq} and formulate the direct Rashba-Edelstein effect in Sec.~\ref{sec:Edelstein}.
Next, considering the interfacial exchange coupling, the REMR is analytically calculated for the dirty and clean interfaces in Sec.~\ref{sec:dirty} and \ref{sec:clean}, respectively.

\subsection{Boltzmann equation}
\label{sec:BoltzmannEq}

Throughout our calculation, we assume that these spin-orbit interactions are substantially larger than the temperature and the energy broadening due to impurity scattering in the 2DEG.
In the following, we refer to this condition as the weak-impurity condition.
Then, the distribution function of the conduction electrons can be expressed as $f({\bm k},\gamma)$, where ${\bm k}$ and $\gamma$ are the wavenumber and the index of the spin-polarized bands, respectively~\footnote{When considering spin-orbit interactions such as Rashba and Dresselhaus within the general framework of the Boltzmann equation, the non-equilibrium distribution function for 2DEG electrons typically manifests as a $2 \times 2$ matrix. In this study, given that the band splitting from these spin-orbit interactions is substantially greater than the energy broadening due to impurity scattering, only the diagonal components of this $2 \times 2$ matrix are considered in the non-equilibrium distribution function.
For a detailed discussion, see Ref.~\onlinecite{Suzuki2023}.}.
Based on the perturbation theory with respect to $H_{\rm imp}$ and $H_{\rm int}$, the Boltzmann equation is written as
\begin{align}
\frac{eE_{x}}{\hbar} \frac{\partial f(\bm{k},\gamma)}{\partial k_{x}}
=\frac{\partial f(\bm{k},\gamma)}{\partial t} \Bigl{|}_{\rm imp}
+\frac{\partial f(\bm{k},\gamma)}{\partial t} \Bigl{|}_{\rm int},
\label{eq:BolzmannEquation}
\end{align}
where $e$ ($<0$) is the electron charge.
The first and second terms on the right-hand side describe collision terms due to impurity scattering and interfacial exchange coupling, respectively.
The explicit forms of these collision terms are given in Appendix~\ref{app:CollisionTerms}.

Using the solution of this Boltzmann equation for $f({\bm k},\gamma)$, the spin and current densities in 2DEG are described as
\begin{align}
{\bm s}
&=\frac{\hbar}{2\mathcal{A}}\sum_{\bm{k},\gamma}
\langle \bm{k}\gamma|\hat{\bm{\sigma}}|\bm{k}\gamma\rangle f(\bm{k},\gamma) ,
\label{eq:defs} \\
\bm{j} &=\frac{e}{\mathcal{A}}\sum_{\bm{k},\gamma} \bm{v}(\bm{k},\gamma) f(\bm{k},\gamma), \label{jREMR}
\end{align}
where $\bm{v}$ is electron velocity defined by
\begin{align}
\bm{v}(\bm{k},\gamma)&=
\frac{1}{\hbar}\frac{\partial E^{\gamma}_{\bm{k}}}{\partial\bm{k}}
=\frac{\hbar\bm{k}}{m^*}+\frac{\gamma}{\hbar}\frac{\partial h_{\rm eff}(\bm{k})}{\partial\bm{k}} .
\label{eq:defv}
\end{align}
In the following calculation, the summation with respect to ${\bm k}$ is replaced with an integral as
\begin{align}
\frac{1}{\mathcal{A}}\sum_{\bm{k}} (\cdots) = \frac{1}{2\pi} \int_{0}^{\infty}dk\, |\bm k| \int_{0}^{2\pi}\frac{d\varphi} {2\pi}(\cdots). \label{eq:sumint}
\end{align}

\subsection{Direct Rashba-Edelstein effect}
\label{sec:Edelstein}

Next, we briefly explain how to describe the direct Rashba-Edelstein effect for 2DEG under a DC electric field $\bm{E}=(E_{x},0)$.
For this purpose, we omit the collision term due to the interface in the Boltzmann equation (\ref{eq:BolzmannEquation}).
The distribution function is described by the following form
\begin{align}
f({\bm k},\gamma) &= f_0(E_{\bm k}^\gamma) + f_1({\bm k},\gamma), \label{eq:f1def} \\
f_1({\bm k},\gamma) &= -\frac{\partial f_0(E_{\bm k}^\gamma)}{\partial E_{\bm k}^\gamma} \delta \mu_1(\bm k,\gamma),
\label{eq:f1org}
\end{align}
where $f_{0}(\epsilon)=(\exp[\upbeta(\epsilon-\mu)]+1)^{-1}$ is the Fermi distribution function, $\upbeta$ is the inverse temperature, $f_1({\bm k},\gamma)$ describes a modulation by the external electric field, and $\delta \mu_1(\bm{k},\gamma)$
denotes a chemical potential shift.
Within the linear response to the electric field, $\delta \mu_1(\bm k,\gamma)$ is proportional to $E_x$.
By substituting Eqs.~(\ref{eq:f1def}) and (\ref{eq:f1org}) into Eq.~(\ref{eq:BolzmannEquation}) and by picking up the term of linear order of $E_x$ in both sides of Eq.~(\ref{eq:BolzmannEquation}), we obtain the following relation:
\begin{align}
\frac{eE_{x}}{\hbar} \frac{\partial f_0(E^{\gamma}_{\bm k})}{\partial k_{x}}
=\frac{\partial f_1(\bm{k},\gamma)}{\partial t} \Bigl{|}_{\rm imp}.
\label{eq:REequation}
\end{align}
By straightforward calculation of the Boltzmann equation, the chemical potential shift is finally obtained as
\begin{align}
\delta \mu_1({\bm k},\gamma)
&=\frac{\hbar^{2} eE_{x}|\bm{k}|}{\Gamma m^{*}} \cos\varphi,\label{eq:fsolimp1}
\end{align}
where $\Gamma=2\pi n_{\rm imp}u^{2}D(\epsilon_{\rm F})$ is the energy broadening due to impurities.
For a detailed calculation, see Appendix~\ref{app1}. 
It should be noted that this result is consistent with that of Ref.~\cite{Trushin2007}.

\subsection{Dirty interface}
\label{sec:dirty}

Next, we consider the Rashba-Edelstein magnetoresistance (REMR) at a dirty FI/2DEG interface, for which the matrix element of the interfacial exchange coupling is momentum independent as given in Eq.~(\ref{eq:dirty}).
We consider the nonequilibrium distribution function in the following form:
\begin{align}
f(\bm{k},\gamma)
&= f_{0}(E^{\gamma}_{\bm k})+f_1(\bm{k},\gamma)+f_{\rm D}(\bm{k},\gamma), \label{eq:fdirty1} \\
f_{\rm D}(\bm{k},\gamma)&=-\frac{\partial f_{0}(E^{\gamma}_{\bm k})}{\partial E^{\gamma}_{\bm{k}}} \delta\mu_{\rm D}({\bm k},\gamma),
\label{eq:fdirty2}
\end{align}
where $f_{\rm D}(\bm{k},\gamma)$ denotes a modulation of the distribution function due to the interfacial exchange interaction and $f_1(\bm{k},\gamma)$ is given by Eqs.~(\ref{eq:f1org}) and (\ref{eq:fsolimp1}).
Considering the second-order perturbation with respect to the interfacial Hamiltonian, the collision term due to the interfacial scattering becomes proportional to ${\rm max}(|\bar{T}|^{2},|\bar{\mathcal{T}}|^{2})$ through the transition rates.
Since the distribution function contributing to the REMR is proportional to ${\rm max}(|\bar{T}|^{2}E_{x},|\bar{\mathcal{T}}|^{2}E_{x})$, we evaluate the chemical potential shift $\delta\mu_{\rm D}({\bm k},\gamma)$ up to this order.
We note $|\delta\mu_{\rm D}({\bm k},\gamma)| \ll |\delta\mu_1({\bm k},\gamma)|$.
By substituting Eqs.~(\ref{eq:fdirty1})-(\ref{eq:fdirty2}) into the Boltzmann equation (\ref{eq:BolzmannEquation}) and by comparing the terms of order of ${\rm max}(|\bar{T}|^{2}E_{x},|\bar{\mathcal{T}}|^{2}E_{x})$ in both sides, we obtain
\begin{align}
0=\left.\frac{\partial f_{\rm D}(\bm{k},\gamma)}{\partial t} \right|_{\rm imp}
+\left. \frac{\partial f_1(\bm{k},\gamma)}{\partial t} \right|_{\rm int}.\label{eq:OET2}
\end{align}
After lengthy calculation, the full solution of the Boltzmann equation gives
\begin{align}
\delta \mu_{\rm D} (\varphi,\gamma)&=
\gamma
\frac{2\pi D(\epsilon_{\rm F})S_{0} eE_{x}\mathcal{A}}{\Gamma^{2}} I(T) g(\theta,\varphi), \label{eq:mu2result} \\
I(T)&=-4|\bar{T}|^{2}
\sum_{\bm{q}}\langle N_{\bm{q}}\rangle\sin^2(q_z a)
+S_{0}|\mathcal{\bar{T}}|^{2}, 
\label{eq:I2def} \\
g(\theta,\varphi) &= 
\{ [\alpha+\beta\eta]\sin(\varphi-\theta)+[\beta+\alpha\eta]\cos(\varphi+\theta)\}
\nonumber \\
&\times \frac{\alpha\sin\theta-\beta\cos\theta}{(1-\eta^2)\sqrt{\alpha^2+\beta^2+2\alpha\beta\sin2\varphi}}, \label{eq:g2def}
\end{align}
where $I(T)$ denotes a temperature-dependent factor,  $g(\theta,\varphi)$ is an angle-dependent factor,
$N_{\rm FI}$ is the number of unit cells in the FI, and $\eta$ is a factor defined as
\begin{align}
\eta
& =\begin{cases}
\beta/\alpha & (\alpha^{2} \geq \beta^{2})\\
\alpha/\beta & (\beta^{2} \geq \alpha^{2})
\end{cases}.\label{def:eta}
\end{align}
Here, we have omitted terms independent of $\theta$ since they do not contribute to the REMR.
For a detailed calculation, see Appendix~\ref{app:dirty}.

The REMR is described by $f_{\rm D}({\bm k},\gamma)$, which is a modulation due to the FI/2DEG interface.
The modulation of the current density is given as
\begin{align}
\Delta\bm{j}_{\rm D}(\theta)
&=\frac{e}{2\pi}\sum_{\gamma}
\int_{0}^{\infty}\! \! dk \, |\bm k|\int_{0}^{2\pi}\! \frac{d\varphi}{2\pi} \bm{v}(\bm{k},\gamma) f_{\rm D}(\bm{k},\gamma). 
\label{jREMR1}
\end{align}
Using the solution of the Boltzmann equation given in Eqs.~(\ref{eq:mu2result})-(\ref{eq:g2def}), the current modulation is calculated as
\begin{align}
\Delta\bm{j}_{\rm D}(\theta)& =\frac{e^{2}k_{\rm F}D(\epsilon_{\rm F})S_{0}E_{x}\mathcal{A}I(T)}{\hbar^{2} v_{\rm F}\Gamma^{2}} \nonumber \\
&\times \left( \begin{array}{c}
(\alpha\sin\theta-\beta\cos\theta)^{2} \\
-(\alpha^{2}+\beta^{2})\cos\theta\sin\theta \end{array} \right), \label{eq:REMRendxynew}
\end{align}
where $v_{\rm F}=\hbar k_{\rm F}/m^{*}$ is the Fermi velocity in the absence of the spin-orbit interactions.
In a similar way, the modulation of the spin density is calculated as
\begin{align}
\Delta\bm{s}_{\rm D}(\theta)
&=\frac{\hbar}{4\pi}\sum_{\gamma}\int \! dk \, |{\bm k}| \! \int \frac{d\varphi}{2\pi} 
\langle \bm{k}\gamma|\hat{\bm{\sigma}}|\bm{k}\gamma\rangle f_{\rm D}(\bm{k},\gamma) \nonumber \\
&\hspace{-5mm}=
\frac{k_{\rm F}D(\epsilon_{\rm F})S_{0} eE_{x}\mathcal{A}I(T)}{2v_{\rm F}\Gamma^{2}} \nonumber \\
&\hspace{-5mm} \times
\frac{\alpha\sin\theta-\beta\cos\theta}{1-\eta^{2}}
\begin{pmatrix}
\cos\theta & \sin\theta \\
\sin\theta & \cos\theta
\end{pmatrix}
\begin{pmatrix}
1+\eta^{2} \\ -2\eta
\end{pmatrix}.\label{eq:sREMRend}
\end{align}
For detailed calculation, see Appendix~\ref{app:dirty}.

\subsection{Clean interface}
\label{sec:clean}

Next, we consider the REMR for a clean interface, using the condition given in Eq.~(\ref{eq:clean}).
The nonequilibrium distribution function of the 2DEG electrons can be expressed in the following form:
\begin{align}
f(\bm{k},\gamma)
&\equiv f_{0}(\bm{k},\gamma)+f_1(\bm{k},\gamma)+f_{\rm C}(\bm{k},\gamma),
\label{eq:fclean1} \\
f_{\rm C}(\bm{k},\gamma)
&=-\frac{\partial f_{0}(E^{\gamma}_{\bm k})}{\partial E^{\gamma}_{\bm{k}}} \delta\mu_{\rm C}({\bm k},\gamma).
\label{eq:fclean2} 
\end{align}
This form is the same as Eqs.~(\ref{eq:fdirty1}) and (\ref{eq:fdirty2}) except for the subscription `C', which indicates the case of the clean interface.
We note that $\delta\mu_{\rm C}({\bm k},\gamma)$ is of order of ${\rm max}(E_{x}|\bar{T}|^{2},E_{x}|\bar{\mathcal{T}}|^{2})$ and $|\delta\mu_{\rm C}({\bm k},\gamma)|\ll |\delta\mu_1({\bm k},\gamma)|$.
By substiting Eqs.~(\ref{eq:fclean1})-(\ref{eq:fclean2}) into the Boltzmann equaiton (\ref{eq:BolzmannEquation}) and by comparing the terms of order of ${\rm max}(|\bar{T}|^{2}E_{x},|\bar{\mathcal{T}}|^{2}E_{x})$ in both sides, we obtain
\begin{align}
0=\frac{\partial f_{\rm C}(\bm{k},\gamma)}{\partial t} \Bigl{|}_{\rm imp}
+\frac{\partial f_1(\bm{k},\gamma)}{\partial t} \Bigl{|}_{\rm int}.\label{eq:OET2C}
\end{align}
After lengthy calculation, the spin and current densities are analytically obtained as
\begin{align}
\Delta\bm{j}_{\rm C}({\theta}) &= \int_{0}^{2\pi}\frac{d\varphi}{2\pi}
\frac{4 e^{2}m^{*} D(\epsilon_{\rm F})\mathcal{A}S_{0}|\bar{T}|^{2}E_{x}}{\hbar^{3}\Gamma^{2} \kappa(\varphi) } \mathcal{J}(\varphi) \nonumber \\
&\times \left( \begin{array}{c}
(\alpha^{2}+\beta^{2})\cos\varphi+2\alpha\beta\sin 3\varphi \\
(\alpha^{2}+\beta^{2})\sin\varphi-2\alpha\beta\cos3\varphi
\end{array} \right),\label{eq:JbJC}\\
\Delta\bm{s}_{\rm C}(\theta) &=\frac{2k_{\rm F}D(\epsilon_{\rm F})S_{0}|\bar{T}|^{2}\mathcal{A}eE_{x}}{v_{\rm F}\Gamma^{2}}
\int_0^{2\pi} \frac{d\varphi}{2\pi}
\hat{\bm{h}}_{\rm eff}(\varphi)
\mathcal{J}(\varphi), \label{eq:sbJC}
\end{align}
where $\kappa(\varphi)$ is defined in Eq.~(\ref{def:kappa}) and $\mathcal{J}(\varphi)$ is given as
\begin{align}
& \mathcal{J}(\varphi) = B(\varphi,\theta) \nonumber \\ 
& \hspace{4mm} +\int_{0}^{2\pi}\frac{d\varphi''}{2\pi} B(\varphi'',\theta) \hat{\bm{h}}^T_{\rm eff}(\varphi) \hat{M} \hat{\bm{h}}_{\rm eff}(\varphi''), \label{eq:defmathJ}\\
& B(\varphi,\theta) =  
\mathcal{I}_{1}(T) \kappa(\varphi) \cos\varphi [\hat{\bm{h}}_{\rm eff}(\varphi)\cdot\hat{\bm{m}}(\theta)]^2
\nonumber \\ &\hspace{4mm} -2\mathcal{I}_{2}(T) \sin\varphi[\hat{\bm{h}}_{\rm eff}(\varphi)\cdot\hat{\bm{m}}(\theta)]
[\bm{g}(\varphi)
\cdot\hat{\bm{m}}(\theta)], \\
&\hat{M}=\frac{2}{1-\eta^2} \left(\begin{array}{cc} 1 & -\eta \\ -\eta & 1 \end{array}\right), \\
&\bm{g}(\varphi) =
\begin{pmatrix}
\alpha\cos\varphi
-\beta\sin\varphi
\\
\alpha\sin\varphi
-\beta\cos\varphi
\end{pmatrix} \label{eq:bmgdef},\\
& \mathcal{I}_{1}(T) =
\sum_{q_{z}>0}\int_{0}^{2\pi}\frac{d\varphi}{2\pi}
N(\varphi,q_{z})\sin^2(q_z a)
\cos\varphi[1-2\cos\varphi],\label{def:I1}\\
& \mathcal{I}_{2}(T) =
\sum_{q_{z}>0}\int_{0}^{2\pi}\frac{d\varphi}{2\pi}
N(\varphi,q_{z})\sin^2(q_z a)
\sin^{2}\varphi , \label{def:I2}\\
&N(\varphi,q_{z}) 
=\frac{1}{e^{\beta\hbar\omega(\varphi,q_{z})}-1},\label{eq:hoNdef}\\
&\hbar\omega(\varphi,q_{z})
= \hbar|\gamma_{\rm g}|h_{\rm dc}+4\mathcal{D}k_{\rm F}^{2}\sin^{2} \frac{\varphi}{2} +\mathcal{D}q_{z}^{2}. \label{def:hokkq} 
\end{align}
Here, we have omitted terms independent of $\theta$ since they do not contribute to the REMR.
For a detailed calculation, see Appendix~\ref{app:clean}.

\section{Result}
\label{sec:results}

In this section, we show our results for the REMR.
First, we briefly discuss the effect of interface randomness in Sec.~\ref{sec:resultQF}.
Next, Secs.~\ref{sec:ResultRashba} and \ref{sec:Resultaob11} present the results for the cases where the 2DEG has only Rashba spin-orbit interaction ($\beta = 0$) and where Rashba and Dresselhaus spin-orbit interactions counterbalance each other ($\alpha = 1.1\beta$), respectively.
Finally, we discuss their maximum values as a function of $\alpha/\beta$ in Sec.~\ref{sec:MaxIamp}.

In this section, we express the constant prefactors for the spin density and charge current density for a dirty interface as  
\begin{align}
s_{x,{\rm D}}
&= -\frac{k_{\rm F}D(\epsilon_{\rm F})S_{0}^{2} eE_{x}\mathcal{A}|\bar{\mathcal{T}}|^{2} x}{2v_{\rm F}\Gamma^{2}}, \quad (x=\alpha,\beta), \label{eq:defsTMa}\\
j_{x,{\rm D}}
&= \frac{e^{2}k_{\rm F}D(\epsilon_{\rm F})S_{0}^{2}E_{x}\mathcal{A}|\bar{\mathcal{T}}|^{2}x^{2}}{\hbar^{2} v_{\rm F}\Gamma^{2}}, \quad (x=\alpha,\beta). \label{eq:defjTMa}
\end{align}  
Similarly, for a clean interface, we define  
\begin{align}
s_{x,{\rm C}}
&= -\frac{2k_{\rm F}^{2}LD(\epsilon_{\rm F})S_{0}|\bar{T}|^{2}\mathcal{A}eE_{x}x}{\pi v_{\rm F}\Gamma^{2}}, \quad (x=\alpha,\beta), \label{def:tilsa}\\
j_{x,{\rm C}}
&= \frac{4k_{\rm F}L e^{2}m^{*} D(\epsilon_{\rm F})\mathcal{A}S_{0}|\bar{T}|^{2}E_{x}x^{2}}{\pi\hbar^{3}\Gamma^{2}}, \quad (x=\alpha,\beta). \label{def:tilJa}
\end{align}  
Here, $L$ denotes the thickness of the FI. Note that all these constants are positive for $E_{x} > 0$. Furthermore, the normalization factors for the charge current, $j_{x,{\rm D}}$ and $j_{x,{\rm C}}$, are proportional to $\alpha^2$ or $\beta^2$. This scaling behavior is reasonable, as the charge current arises from the combined effect of the direct and inverse Rashba-Edelstein effects, both of which are induced by spin-orbit interaction.  

In the following, we plot the dimensionless spin and charge current densities defined as $\Delta {\bm s}_{\rm D}/s_{\alpha,{\rm D}}$ and $\Delta {\bm j}_{\rm D}/j_{\alpha,{\rm D}}$ (or $\Delta {\bm s}_{\rm D}/s_{\beta,{\rm D}}$ and $\Delta {\bm j}_{\rm D}/j_{\beta,{\rm D}}$) for the dirty interface, and $\Delta {\bm s}_{\rm C}/s_{\alpha,{\rm C}}$ and $\Delta {\bm j}_{\rm C}/j_{\alpha,{\rm C}}$ (or $\Delta {\bm s}_{\rm C}/s_{\beta,{\rm C}}$ and $\Delta {\bm j}_{\rm C}/j_{\beta,{\rm C}}$) for the clean interface.

\subsection{Effect of interface randomness}
\label{sec:resultQF}

We first discuss the effect of the interface randomness by comparing the results for the dirty and clean interfaces.
Let us start with a discussion on the factor $I(T)$ for the dirty interface, which is given in Eq.~(\ref{eq:I2def}).
We note that the second term of $I(T)$ describes the exchange bias proportional to $|\bar{\cal T}|^2$, which originates from the second term of the interfacial Hamiltonian, Eq.~(\ref{eq:Hintdirty}).
Moreover, $|\bar{\cal T}|^2$ is larger than $|\bar{T}|^2$ by approximately the large factor $N_{\rm FI}$, and thus the second term of $I(T)$ in Eq.~(\ref{eq:I2def}) is dominant (see Appendix~\ref{app:DrvHint} for a detail). 
This means that the REMR is mainly induced by the exchange bias term for the dirty interface.
In the following discussion, we approximate
\begin{align}
I(T) \simeq S_0  |\bar{\cal T}|^2, \label{eq: defITSNT}
\end{align}
for simplicity.
We stress that this approximation predicts a temperature-independent REMR for the dirty interface.

In contrast, the coupling strength $\bar{\cal T}$ due to the exchange bias does not appear in the analytical results for the clean interface, which are given in Eqs.~(\ref{eq:JbJC})-(\ref{eq:bmgdef}).
This indicates that for the clean interface the REMR is induced not by exchange bias but by dynamic magnon absorption(emission) processes, which is described by the first term in Eq.~(\ref{eq:Hd+Hs}).
As a result, the REMR is temperature dependent.
This difference between the dirty and clean interfaces is one of our main results.

\subsection{Rashba spin-orbit interaction ($\beta=0$)}
\label{sec:ResultRashba}

\begin{figure*}[tbp]
\centering
\includegraphics[width=155mm]{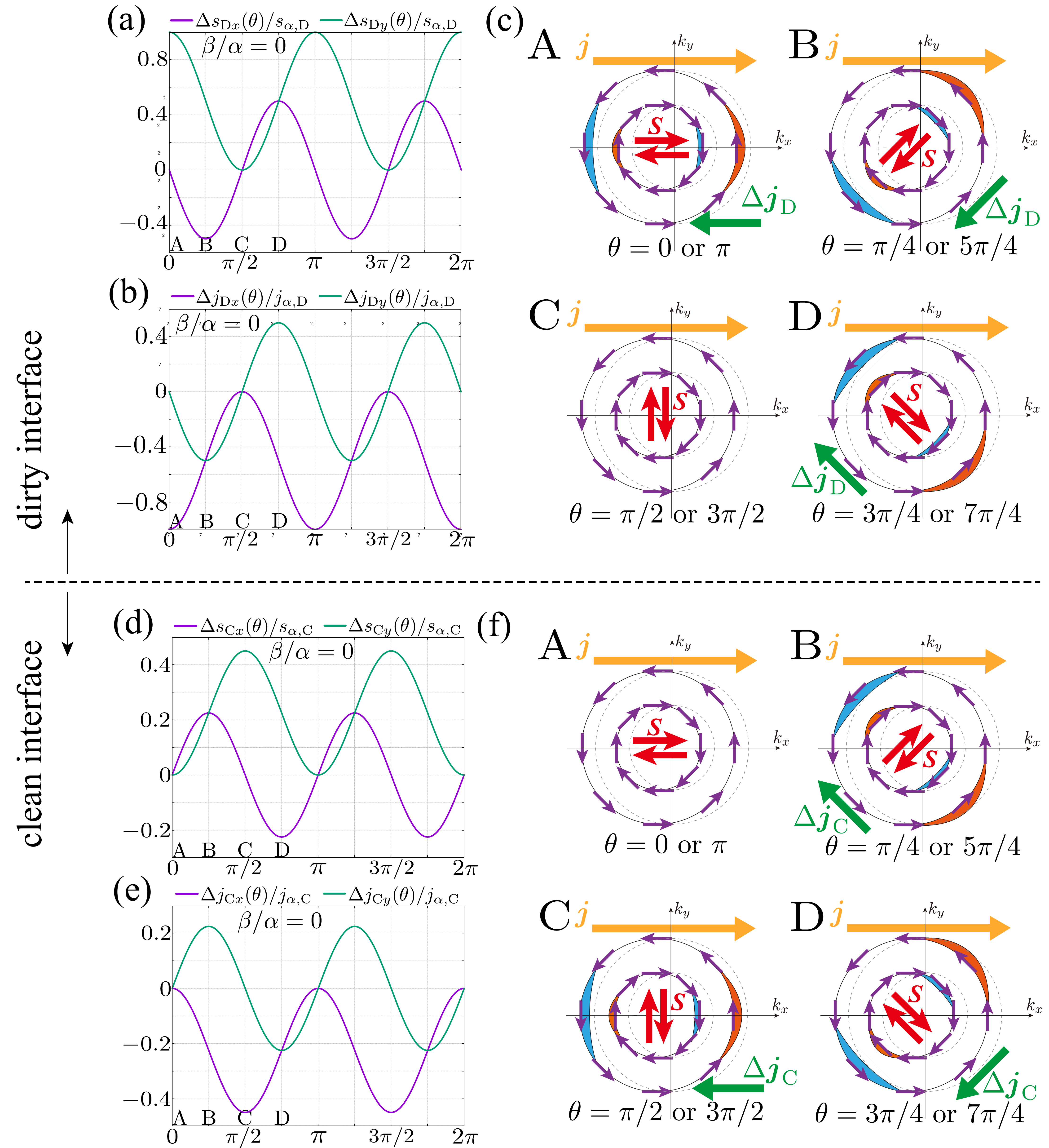}
\caption{Modulations of spin and charge current densities induced by the REMR as a function of the spin azimuth angle $\theta$ of the FI in the presence of only Rashba spin-orbit interaction ($\beta=0$). (a), (b) Modulation of spin density and charge current density at a dirty interface. For clarity, the modulations are set to zero at the reference points $\theta = \pi/2, 3\pi/2$. (c) Schematic illustration of the modulation of nonequilibrium distribution functions at a dirty interface for (A) $\theta=0,\pi$, (B) $\theta=\pi/4,5\pi/4$, (C) $\theta=\pi/2,3\pi/2$, and (D) $\theta=3\pi/4,7\pi/4$. The orange (blue) regions indicate an increase (decrease) in the distribution function of conduction electrons relative to the reference points $\theta=\pi/2,3\pi/2$. The red and green arrows denote the directions of spin polarization in the FI and the charge current modulation in the 2DEG, respectively.  
(d), (e) Modulations of spin density and charge current densities at a clean interface.  Note that the modulations are set to zero at the reference points $\theta = 0,\pi$, for simplicity of explanation. (f) Schematic illustration of the modulation of nonequilibrium distribution functions at a clean interface. For a clean interface, the parameters are set as $k_{\rm B}T/\hbar\omega_{\bm 0}=3$, $|\gamma_{\rm g}|h_{\rm dc}/\omega_{\bm 0}=0.1$, and $k_{\rm F}a=0.1$, where $\hbar\omega_{\bm 0}=4\mathcal{D}k_{\rm F}^2$ and $a$ is the lattice constant of the FI.}
\label{fig:Rashba}
\end{figure*}

In this section, we discuss the case where only the Rashba spin-orbit interaction exists ($\beta = 0$). Figure~\ref{fig:Rashba} illustrates the spin and charge current densities in the 2DEG as a function of the spin orientation $\theta$ of the FI.  
We first present the results for the dirty interface in Sec.~\ref{sec:ResultDirty},
and next present the results for the clean interface in Sec.~\ref{sec:ResultClean}.

\subsubsection{Dirty interface}
\label{sec:ResultDirty}

In this section, we discuss the results of Figs.~\ref{fig:Rashba}(a) and (b) that plot the modulation of the spin density and charge current density in the 2DEG induced by the REMR at the dirty interface. Because only the relative modulation induced by a change of $\theta$ is relevant to REMR, we set the origin of the modulation at $\theta=\pi/2$, that is, set $\Delta {\bm s}_{\rm D}(\pi/2)$ and $\Delta {\bm j}_{\rm D}(\pi/2)$ as zero.
Both spin and current densities are periodic functions of $\theta$ with a period $\pi$.
We note that the direction of the spin and current densities rotates as $\theta$ changes.

This result is intuitively explained as follows.
The four pictures of Fig.~\ref{fig:Rashba}(c) show schematic diagrams of the spin-splitting Fermi surface and the modulation of the distribution function for (A) $\theta = 0,\pi$, (B) $\theta = \pi/4,5\pi/4$, (C) $\theta =\pi/2,3\pi/2$, and (D) $\theta = 3\pi/4,7\pi/4$, respectively.
Here, we set the modulations of the spin and current densities as zero at (C) $\theta=\pi/2,3\pi/2$ because we are interested only in the relative modulation measured from the reference points.
The dashed lines in these diagrams represent the equilibrium position of the Fermi surface in the absence of an external DC electric field. 
When an external DC electric field is applied in the $+x$ direction, the Fermi surface shifts in the $-x$ direction, resulting in the direct Rashba-Edelstein effect that induces spin accumulation in the $-y$ direction. 

As discussed in Sec.~\ref{sec:resultQF}, the static exchange bias across the junction contributes dominantly to the REMR for the dirty interface.
This exchange bias acts on conduction electrons as a static Zeeman field and causes spin relaxation of conduction electrons near the Fermi surface.
We note that spin relaxation is enhanced when the spin polarization axis of conduction electrons is perpendicular to this effective Zeeman field.
As an example, let us consider the case of $\theta=0,\pi$, whose distribution function is schematically shown in the diagram A of Fig.~\ref{fig:Rashba}(c).
In this case, spin flipping of the conduction electrons is caused at the place where the spin polarization of the Fermi surface is perpendicular to ${\bm S}$, i.e., in the $\pm y$ direction.
As a result, the distribution function of the conduction electrons with spins oriented in the $-y$ direction is reduced and that in the $+y$ direction is enhanced.
The relative change of the distribution function at (A) $\theta=0,\pi$ from (C) $\theta=\pi/2,3\pi/2$ is indicated by orange and blue regions in the diagram A of Fig.~\ref{fig:Rashba}(c).
Furthermore, this change in spin accumulation causes the inverse Rashba-Edelstein effect, generating a current modulation $\Delta\bm{j}_{\rm D}$ in the $-x$ direction. 

A similar explanation is possible for (B) $\theta=\pi/4, 5\pi/4$  and (D) $\theta=3\pi/4, 7\pi/4$.
As an example, let us consider the case of (B) $\theta=\pi/4, 5\pi/4$.
In this case, spin flipping is caused by exchange bias, where the spin polarization of the Fermi surface is in the $3\pi/4$ and $7\pi/4$ directions perpendicular to the spin ${\bm S}$ of the FI.
As a result, the distribution function of the conduction electrons with spins oriented in the $7\pi/4$ ($3\pi/4$) direction is reduced (enhanced), leading to current modulation $\Delta\bm{j}_{\rm D}$ in the $5\pi/4$ direction.

\subsubsection{Clean interface}
\label{sec:ResultClean}

In this section, we discuss the results of Figs.~\ref{fig:Rashba}(d) and (e) that plot the modulation of the spin density and charge current density in the 2DEG induced by the REMR at the clean interface.
These plots illustrate how the spin and current densities in the 2DEG depend on the spin angle $\theta$ of the FI, where we set the origin of the modulation at $\theta=0$, that it, set $\Delta {\bm s}_{\rm C}(\theta=0)=0$ and $\Delta {\bm j}_{\rm C}(\theta=0)=0$.
We find that the direction of the spin and current densities rotates as $\theta$ changes.
Here, we should note that the sign of the REMR is opposite to that in the dirty interface; the spin and current densities for the clean interface take maximum (minimum) values when they take minimum (maximum) values for the dirty interface (compare Figs.~\ref{fig:Rashba}(d) and (e) with Figs.~\ref{fig:Rashba}(a) and (b)).

This contrast result is explained by the difference of the physical process in the interfacial spin-flipping scattering.
For the clean interface, the effective Zeeman field due to the exchange bias is not effective, and the dynamic process by the magnon absorption/emission is dominant in the REMR.
As a result, spin flipping of conduction electrons near the Fermi surface is caused by spin transfer due to magnon absorption or emission, which carries a spin in the direction of $-{\bm S}$, and therefore is enhanced when the spin polarization axis of conduction electrons is {\it parallel} to the spin ${\bm S}$ in the FI.
This difference in the spin-flipping process of conduction electrons shifts the dependence on $\theta$ by $\pi/2$ compared to the case of the dirty interface.

The remaining explanation is common as the dirty interface except for the direction of the spin relaxation (see the four right diagrams of Fig.~\ref{fig:Rashba}(f)).
For example, let us consider the case of $\theta=\pi/2,3\pi/2$, whose distribution function is schematically shown in the diagram C of Fig.~\ref{fig:Rashba}(f).
Spin flipping of conduction electrons occurs where the spin polarization of the Fermi surface is parallel to ${\bm S}$, that is, in the $\pm y$ direction.
As a result, the distribution function of the conduction electrons with spins oriented in the $-y$ ($+y$) direction is reduced (enhanced), as indicated by the orange and blue regions in the diagram C of Fig.~\ref{fig:Rashba}(f).
We note that this change of the distribution function reduces the spin accumulation driven by the static electric field.
This change in the distribution function generates a current modulation $\Delta\bm{j}_{\rm C}$ in the $-x$ direction.
The behaviors of the spin and current densities at (B) $\theta=\pi/4, 5\pi/4$ and (D) $\theta=3\pi/4, 7\pi/4$ can also be explained in a similar way.

\subsection{Competing spin-orbit interactions ($\alpha/\beta=1.1$)}
\label{sec:Resultaob11}

\begin{figure*}[tbp]
\centering
\includegraphics[width=130mm]{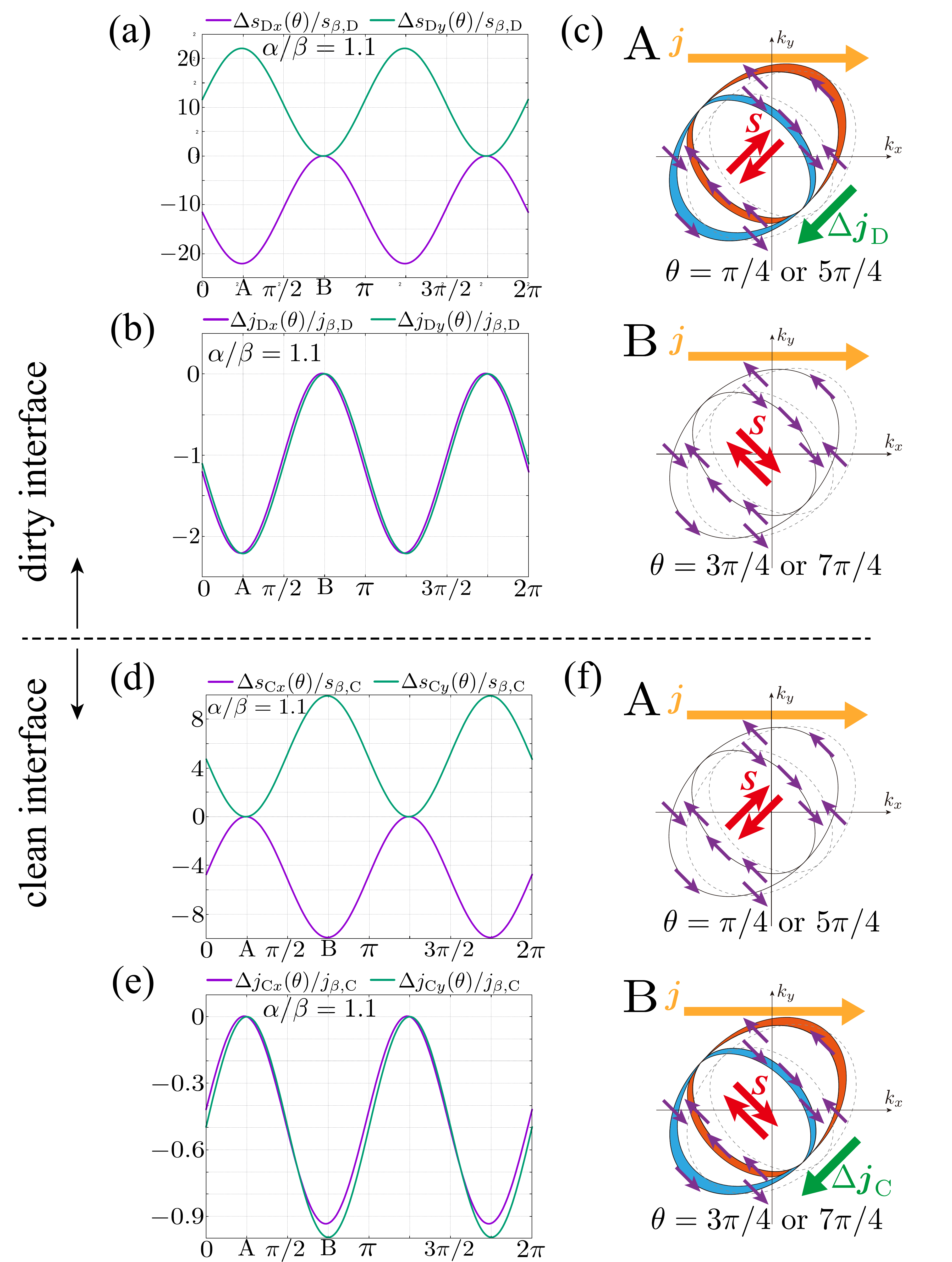}
\caption{Modulations of spin and charge current densities induced by the REMR as a function of the spin azimuth angle $\theta$ of the FI for $\alpha/\beta=1.1$. (a), (b) Modulation of spin density and charge current density at a dirty interface. Note that the modulations are set to zero at the reference points $\theta = 3\pi/4, 7\pi/4$. (c) Schematic illustration of the modulation of nonequilibrium distribution functions at a dirty interface for (A) $\theta=\pi/4,5\pi/4$ and (B) $\theta=3\pi/4,7\pi/4$. 
(d), (e) Modulations of spin density and charge current densities at a clean interface.  Note that the modulations are set to zero at the reference points $\theta = \pi/4,5\pi/4$. (f) Schematic illustration of the modulation of nonequilibrium distribution functions at a clean interface.
For a clean interface, the parameters are set as $k_{\rm B}T/\hbar\omega_{\bm 0}=3$, $|\gamma_{\rm g}|h_{\rm dc}/\omega_{\bm 0}=0.1$, and $k_{\rm F}a=0.1$, where $\hbar\omega_{\bm 0}=4\mathcal{D}k_{\rm F}^2$ and $a$ is the lattice constant of the FI.}
\label{fig:aob11}
\end{figure*}

Next, we consider the case of competing magnitudes of Rashba and Dresselhaus spin-orbit interactions ($\alpha\simeq \beta$).
Figure~\ref{fig:aob11} depicts the spin and charge current densities in the 2DEG as a function of the spin orientation $\theta$ of the FI.
We first present the results for the dirty interface in Sec.~\ref{sec:aob11Dirty},
and next present the results for the clean interface in Sec.~\ref{sec:aob11Clean}.

\subsubsection{Dirty interface}
\label{sec:aob11Dirty}

In this section, we discuss the relative modulations of the spin and charge current densities, measured from $\theta = 3\pi/4$, in the dirty interface shown in Figs.~\ref{fig:aob11}(a) and (b).
Compared to $\theta = 3\pi/4$, the spin modulation $\Delta {\bm s}_{\rm D}$ is induced in the direction of $3\pi/4$ while the current modulation $\Delta {\bm j}_{\rm D}$ is induced in the direction of $5\pi/4$.
At $\theta = \pi/4, 5\pi/4$, both have the maximum relative modulation measured from $\theta = 3\pi/4$.

This result is intuitively explained as follows.
Figure~\ref{fig:aob11}(c) includes two diagrams that schematically indicate the modulation of the distribution function in (A) $\theta = \pi/4,5\pi/4$ and (B) $\theta=3\pi/4,7\pi/4$. 
We note that the modulation is set to zero at the reference points (B) $\theta=3\pi/4,7\pi/4$.
In both diagrams, the external DC electric field in the $+x$ direction first shifts the Fermi surface in the $-x$ direction, leading to a $7\pi/4$ directional spin accumulation due to the direct Rashba-Edelstein effect.
For (A) $\theta = \pi/4,5\pi/4$, the spin relaxation in the direction of $3\pi/4$ and $7\pi/4$ is caused by the effective Zeeman field, which is perpendicular to the direction of ${\bm S}$.
The relative change of the distribution function in (A) $\theta = \pi/4,5\pi/4$ measured from (B) $\theta=3\pi/4,7\pi/4$ is indicated by orange and blue regions in the panel A of Fig.~\ref{fig:aob11}(c).
We note that this change of the distribution function reduces the spin accumulation driven by the static electric field.
Furthermore, this change in the distribution function causes a current modulation $\Delta\bm{j}_{\rm D}$ in the direction of $5\pi/4$, compared to the reference point (B).

\subsubsection{Clean interface}
\label{sec:aob11Clean}

Next, in this section, we consider the spin and charge current densities induced by REMR in the clean interface, as depicted in Figs.~\ref{fig:aob11}(d) and (e).
These plots show the modulation of the spin and current densities measured from $\theta = \pi/4,5\pi/4$.
The change in the distribution function from the reference points, $\theta = \pi/4,5\pi/4$, is indicated by the two diagrams of Fig.~\ref{fig:aob11}(f).
Compared to (A) $\theta = \pi/4,5\pi/4$, the modulation of the spin accumulation, $\Delta {\bm s}_{\rm C}$, for (B) $\theta = 3\pi/4,7\pi/4$ is induced in the direction of $3\pi/4$.
As a result, the current modulation $\Delta {\bm j}_{\rm C}$ in (B) $\theta = 3\pi/4,7\pi/4$ is induced in the direction of $5\pi/4$.
We note that the $\theta$ dependence is shifted by $\pi/2$ in comparison with the case of the dirty interface (compare Figs.~\ref{fig:aob11}(d) and (e) with Figs.~\ref{fig:aob11}(a) and (b)).

\subsection{Dependence on $\alpha/\beta$}
\label{sec:MaxIamp}

\begin{figure}[tb]
  \begin{center}
  \includegraphics[width=60mm]{Fig6.pdf}
     \caption{Amplitudes of $x$-compotent of the spin and current modulations, $\Delta s_{x}^{\rm max-min}$ and $\Delta j_{x}^{\rm max-min}$, for the dirty and clean interfaces as a function of $\alpha/\beta$. They are evaluated by the difference between their maximum and minimum values when the spin polarization azimuth of the FI, $\theta$, is changed. For the clean interface, the parameters are set as $k_{\rm B}T/\hbar\omega_{\bm 0}=3$, $|\gamma_{{\rm g}}|h_{\rm dc}/\omega_{\bm 0}=0.1$, and $k_{\rm F}a=0.1$, where $\hbar\omega_{\bm 0}=4\mathcal{D}k_{\rm F}^2$.}
    \label{fig:Fig7}
  \end{center}
\end{figure}

Finally, we discuss the modulation amplitudes of the spin and current densities, which are evaluated by the difference between their maximum and minimum values when the angle $\theta$ is changed.
Fig.~\ref{fig:Fig7} shows $\Delta s_{x}^{\rm max-min}$ and $\Delta j_{x}^{\rm max-min}$ for the $x$ component of spin and current densities as a function of $\alpha/\beta$. 
Here, $\Delta s_{x}^{\rm max-min}$ and $\Delta j_{x}^{\rm max-min}$ represent the difference between the maximum and minimum values of $\Delta s_{x}$ and $\Delta j_{x}$ obtained by varying $\theta$ and $\omega_{\bm{0}}$ for each value of $\alpha/\beta$. For example, in the case of a dirty interface, they can be expressed as follows:
\begin{align}
&\Delta s_{{\rm D}x}^{\rm max-min}(\alpha/\beta)\nonumber \\
&\equiv \max_{\theta,\omega_{\bm{0}}}[\Delta s_{{\rm D}x}(\theta,\omega_{\bm{0}},\alpha/\beta)]-\min_{\theta,\omega_{\bm{0}}}[\Delta s_{{\rm D}x}(\theta,\omega_{\bm{0}},\alpha/\beta)],\\
&\Delta j_{{\rm D}x}^{\rm max-min}(\alpha/\beta)\nonumber \\
&\equiv \max_{\theta,\omega_{\bm{0}}}[\Delta j_{{\rm D}x}(\theta,\omega_{\bm{0}},\alpha/\beta)]-\min_{\theta,\omega_{\bm{0}}}[\Delta j_{{\rm D}x}(\theta,\omega_{\bm{0}},\alpha/\beta)].
\end{align}
In both dirty and clean cases, the amplitude of the spin density modulation diverges at $\alpha/\beta = 1$, while that of the current density modulation monotonically increases as $\alpha/\beta$ increases.
The divergence of $\Delta s_{x}^{\rm max-min}$ originates from complete suppression of spin relaxation at $\alpha/\beta=1$, where the effective Zeeman field due to the Rashba and Dresselhaus spin-orbit interactions is almost aligned along a uniaxial direction (see Fig.~\ref{fig:aob11}(c) or Fig.~\ref{fig:aob11}(f)). 
We note that electron scattering by nonmagnetic impurities under such a uniaxial Zeeman field never causes spin relaxation, leading to divergence of spin relaxation time.
In contrast, the currents generated by the inverse Rashba-Edelstein effect show no divergence at $\alpha/\beta=1$, because they are determined by the difference in the contribution between the outer and inner Fermi surfaces having opposite directions.

\section{Experimental Relevance}
\label{Sec:Experiment}

In this section, we briefly discuss the relevance of our result to the experiment for a Bi/Ag/CoFeB junction system, which has been conducted by Nakayama et al.~\cite{Nakayama2016}
In this junction system, the strong Rashba spin-orbit interaction near the Bi/Ag interface induces the REMR when an external DC electric field is applied.
The longitudinal resistance varies as a function of $\theta$ with a period $\pi$ and takes maximum (minimum) values when $\theta=0$ ($\theta=\pi/2$),
while the transverse resistance shows a periodic dependence on $\theta$ with a
$\pi/4$ phase shift, compared to the longitudinal resistance, in good agreement with the case of the dirty interface in our study~\footnote{Note that the localized spin in the FI is oriented opposite to the external DC magnetic field applied to the FI, therefore, the $\theta+\pi$ in our study corresponds to $\theta$ in Ref.~\onlinecite{Nakayama2016}.}.
This is reasonable because the Fermi wavelength of the carriers in Ag is so short that the interfacial randomness becomes effective.
Thus, our findings qualitatively replicate the experimental result reported in Ref.~\onlinecite{Nakayama2016}.
In the future, the REMR with the opposite sign may be observed if a junction system with a clean interface at which electron scattering due to interfacial randomness can be neglected is realized.

\section{Summary}
\label{sec:summary}

We have theoretically investigated the Rashba-Edelstein magnetoresistance (REMR) in a junction system composed of a ferromagnetic insulator (FI) and a two-dimensional electron gas (2DEG) where the Rashba and Dresselhaus spin-orbit interactions coexist. 
Using a microscopic Hamiltonian and the Boltzmann equation, we calculated the modulation of current and spin densities in the 2DEG induced by REMR under a DC electric field, assuming that the energy broadening due to impurity scattering is significantly smaller than the energy of the spin-orbit interactions. 
We elucidated how these modulations depend on the orientation of spin polarization in the FI and the ratio between the strengths of the two spin-orbit interactions for both dirty and clean FI/2DEG interfaces. 
In the case of a dirty interface, the effective Zeeman field owing to exchange bias contributes to the REMR, whereas in a clean interface, the dynamic process by the magnon absorption/emission is dominant, leading to an opposite sign of REMR compared to that of the dirty interface.
Additionally, we demonstrated that in both interfaces, as the ratio between Rashba and Dresselhaus spin-orbit interactions approaches unity, the modulation of spin density in the 2DEG diverges, while the modulation of current does not show such a singularity.
These findings improve our understanding of the physical mechanisms underlying REMR and will be helpful for interpretation and comparison with the experimental results.
Our formulation of the REMR can be extended in principle to other systems with complex band structures. We leave such an extended analysis for future problems.

\section*{Acknowledgments}

The authors thank Y. Suzuki, Y. Kato, and M. Kohda for their helpful discussions. In particular, we thank Y. Suzuki for his invaluable comments on the precise formulation of the induced current. M. Y. was supported by JST SPRING (Grant No.~JPMJSP2108) and JSPS KAKENHI Grant No.~JP24KJ0624. M. M. was supported by the National Natural Science Foundation of China (NSFC) under Grant No. 12374126, by the Priority Program of Chinese Academy of Sciences under Grant No.~XDB28000000, and by JSPS KAKENHI for Grants (No.~JP23H01839 and No.~JP24H00322) from MEXT, Japan. T. K. acknowledges the support of the Japan Society for the Promotion of Science (JSPS KAKENHI Grant No.~JP20K03831 and No.~JP24K06951).

\appendix

\section{Derivation of the interfacial Hamiltonian}
\label{app:DrvHint}

In this appendix, we derive the Hamiltonian of the interfacial exchange coupling introduced in Eqs.~(\ref{eq:Hd+Hs})–(\ref{eq:clean}) using a simple interface model.
We consider a clean FI in the shape of a cuboid, employ a quantum Heisenberg model on a cubic lattice, and
determine the expressions of $\bar{T}$ and $\bar{\mathcal{T}}$ in Eqs.~(\ref{tqz0}) and (\ref{tqz}).

\subsection{Magnon wave functions}

Using Eqs.~(\ref{eq:HolS+})-(\ref{eq:HolSxp}), the Hamiltonian of the FI is rewritten with the boson operators as follows:
\begin{align}
H_{\rm FI} 
&= J\sum_{\langle i,j\rangle}S_0(-b_{i}^\dagger b_{i}-b_{j}^\dagger b_{j}+b_{i}^\dagger b_{j}+b_{j}^\dagger b_{i}) \nonumber \\
&+\hbar\gamma_{\rm g}h_{\rm dc}\sum_{i}b^\dagger_{i}b_{i} + {\rm const.},
\end{align}
where $J_{ij}$ is approximated as a constant $J$ only for the nearest-neighbor (n.n.) sites. This Hamiltonian can be rewritten as:
\begin{align}
H_{\rm FI}&=\sum_{ij}b^\dagger_{i}(\hat{h}_{\rm 3D})_{ij}b_{j},\label{eq:HFIwa}\\
(\hat{h}_{\rm 3D})_{ij}&=\left\{ \begin{array}{ll} -6JS_{0}+\hbar\gamma_{\rm g}h_{\rm dc}, & (i=j), \\
JS_{0}, & (i\text{ and } j \text{ are n.n.}),\\
0, & ({\rm otherwise}). \end{array} \right.
\end{align}
The eigenenergies and eigenwavefunctions of magnons are obtained by solving the eigenvalue equation:
\begin{align}
\sum_{j}(\hat{h}_{\rm 3D})_{ij}\psi_{\bm{n}}(\bm{r}_{j})=E_{\bm{n}}\psi_{\bm{n}}(\bm{r}_{i}),\label{eq:eigenEmag}
\end{align}
under periodic boundary conditions in the $x$ and $y$ directions and fixed boundary condition in the $z$ direction. Assuming that the interface of the FI-2DEG junction is located at $z=0$, with the number of unit cells in the $x$, $y$, and $z$ directions being $N_{x}$, $N_{y}$, and $N_{z}$, respectively, and the lattice constant of the FI denoted as $a$, the boundary conditions are given by
\begin{align}
& \psi(x+N_{x}a,y,z)=\psi(x,y+N_{y}a,z)=\psi(x,y,z), \\
& \psi(x,y,0)=\psi(x,y,(N_{z}+1)a)=0.
\end{align}
By solving Eq.~(\ref{eq:eigenEmag}), we obtain the eigenvalues and wave functions of magnons as:
\begin{align}
E_{\bm{n}}&=2|J|S_{0}(3-\cos k_{x}a-\cos k_{y}a-\cos k_{z}a)\notag \\
& +\hbar\gamma_{\rm g}h_{\rm dc},\label{eq:E3D}\\
\psi_{\bm{n}}(\bm{r}_{i})&=\sqrt{\frac{2}{N_x N_y (N_z+1)}}e^{ik_{x}x_{i}+ik_{x}x_{y}}
\sin(k_{z}z_{i}).\label{psiNNNzp1sin}
\end{align}
Here, using $\bm{n}=(n_{x},n_{y},n_{z})$, the magnon wave vectors are expressed as:
\begin{align}
k_{x} &= \frac{2\pi n_{x}}{N_{x}a}, \label{eq:kx}\\
k_{y} &= \frac{2\pi n_{y}}{N_{y}a}, \\
k_{z}&=\frac{n_{z}\pi}{(N_{z}+1)a}, \label{eq:kxyznxyz}
\end{align}
where $n_x$ and $n_y$ are integers and $n_z$ is a positive integer.

\subsection{Site representation of $H_{\rm int}$}

The temperature dependence of the REMR is determined by a detail of the interfacial coupling between a metal and a ferromagnetic insulator (FI).
Here, we consider a simple model for it without a lattice mismatch, assuming that the interface is completely flat and the metal and the FI have the same lattice constants.
We consider the interfacial exchange coupling between the site at ${\bm r}_j = (x_j,y_j,a)$ in the FI and the site at ${\bm r}_j' = {\bm r}_j - a \hat{\bm{z}} = (x_j,y_j,0)$ in the 2DEG as
\begin{align}
H_{\rm int} &=\sum_{j}2T_j \bm{S}_{j}\cdot\bm{s}_{j}
=H_{\rm int,d}+H_{\rm int,s}, \\
H_{\rm int,d} &= \sum_{j} T_{j} (S^{x'-}_{j} s^{x'+}_{j} + 
S^{x'+}_{j} s^{x'-}_{j}),  \\
H_{\rm int,s} &= \sum_{j} 2 T_{j} S^{x'}_{j} s^{x'}_{j}, 
\end{align}
where $2T_{j}$ is a strength of the exchange coupling at the bond $j$.
By using the spin-wave approximation, the Hamiltonians are modified as
\begin{align}
&H_{\rm int,d} \simeq
\sqrt{2S_{0}}\sum_{j} T_j (b^{\dagger}_{j}s^{x'+}_{j}+b_{j} s^{x'-}_{j}),\label{eq:Hintdsite}\\
&H_{\rm int,s} \simeq \sum_{j}2T_{j}S_{0}s^{x'}_{j}.\label{eq:Hintssite}
\end{align}
The interfacial randomness is modeled by assuming that $T_j$ is a random variable whose average and variance are given as
\begin{align}
&\langle T_{j} \rangle_{\rm ave} = T_{1}, \label{eq:TjdelT0} \\
&\langle \delta T_{j} \delta T_{j'} \rangle_{\rm ave} = T_{2}^2 \delta_{j,j'} \label{eq:TjdelT},
\end{align}
where $\langle \cdots \rangle_{\rm ave}$ indicates sample average with respect to the interfacial randomness and $\delta T_{j} = T_{j} - \langle T_{j} \rangle_{\rm ave}$.
Let $\bm{R}_{j}=(x_{j},y_{j})$ denote the in-plane coordinates of the interfacial site, and $N_{\rm b}$ the total number of bonds at the FI-2DEG interface. Then, we can write:
\begin{align}
&b_{j}=\sum_{\bm{n}}\psi_{\bm{n}}(\bm{R}_{j},a)b_{\bm{k}(\bm{n})},\label{eq:bjeqpbn}\\
&s^{x'-}_{j}=\frac{1}{N_{\rm b}}\sum_{\bar{\bm{q}}}e^{-i\bar{\bm{q}}\cdot\bm{R}_{j}}s^{x'-}_{\bar{\bm{q}}},\\
&s^{x'+}_{j}=(s^{x'-}_{j})^{\dagger}
=\frac{1}{N_{\rm b}}\sum_{\bar{\bm{q}}}e^{i\bar{\bm{q}}\cdot\bm{R}_{j}}s^{x'+}_{\bar{\bm{q}}},\\
& s^{x'}_{j}=\frac{1}{N_{\rm b}}\sum_{\bar{\bm{q}}}e^{i\bar{\bm{q}}\cdot\bm{R}_{j}}s^{x'}_{\bar{\bm{q}}},
\end{align}
where $\bm{k}(\bm{n})$ is a three-dimensional magnon wavenumber whose components are given in Eqs.~(\ref{eq:kx})-(\ref{eq:kxyznxyz}) and $\bar{\bm q}$ is a two-dimensional wavenumber of conduction electrons in 2DEG.
We note that the number of 2DEG unit cells is assumed to be equal to the number of interfacial bonds. Using these transformations, Eqs.~(\ref{eq:Hintdsite}) and (\ref{eq:Hintssite}) can be rewritten as:
\begin{align}
H_{\rm int,d}
&=\frac{2\sqrt{S_{0}}}{\sqrt{N_{\rm FI}}N_{\rm b}}
\sum_{j}\sum_{\bm{k}}\sum_{\bar{\bm{q}}}\sin(k_{z}a) \notag \\
& \times \Bigl[T_{j}e^{i(-\bm{k}_{\parallel}\cdot\bm{R}_{j}+\bar{\bm{q}}\cdot\bm{R}_{j})}
b_{\bm{k}}^{\dagger}s^{x'+}_{\bar{\bm{q}}} \notag \\
& \hspace{6mm} +T_{j}e^{-i(-\bm{k}_{\parallel}\cdot\bm{R}_{j}+\bar{\bm{q}}\cdot\bm{R}_{j})}
b_{\bm{k}}s^{x'-}_{\bar{\bm{q}}}\Bigr],\\
H_{\rm int,s}&=\frac{2S_{0}}{N_{\rm b}}\sum_{j}\sum_{\bar{\bm{q}}}T_{j}e^{i\bar{\bm{q}}\cdot\bm{R}_{j}}s^{x'}_{\bar{\bm{q}}}.
\end{align}
Here, we approximated $1/\sqrt{N_{z}+1}\simeq 1/\sqrt{N_{z}}$ for $N_{z}\gg1$. Additionally, $N_{\rm FI}=N_{x}N_{y}N_{z}$ denotes the total number of unit cells in the FI, $\bm{k}_{\parallel}=(k_{x},k_{y})$ represents the in-plane components of the magnon wave vector $\bm{k}=(k_x,k_y,k_z)$, and $\bar{\bm{q}}=(\bar{q}_{x},\bar{q}_{y})$ is the wave vector of the 2DEG electrons.

\begin{widetext}

In our study, the interfacial exchange coupling is used to calculate the collision term in the Boltzmann equation (see also Appendix~\ref{app:CollisionTerms}).
The transition rate is determined by the square of the matrix elements, $|\braket{{\bm k}'\gamma';\{N_{\bm q}'\}| H_{\rm int,d}| {\bm k}\gamma;\{N_{\bm q}\}}|^2$ and $|\braket{{\bm k}'\gamma'| H_{\rm int,s}| {\bm k}\gamma}|^2$, where $\ket{{\bm k}\gamma}$ denotes a one-electron state with a wavenumber ${\bm k}$ and a spin state $\gamma$ ($=\pm$) and $\ket{{\bm k}\gamma;\{ N_{\bm q}\}} = \ket{{\bm k} \gamma}\otimes \prod_{\bm q} \ket{N_{\bm q}}$ is a combination with a magnon number state with a wavenumber ${\bm q}$.
By takeing the random average with respect to the interfacial coupling, the square of the matrix elements becomes
\begin{align}
& \Bigl\langle |
\langle \bm{k}'\gamma'; \{ N'_{\bm{q}} \} |
H_{\rm int,d}|\bm{k}\gamma; \{ N_{\bm{q}} \} \rangle
|^2 \Bigr\rangle_{\rm ave} \notag \\
&= \frac{4S_{0}}{N_{\rm FI}N_{\rm b}^2}
\sum_{\bm{k}_{1}} \sum_{\bar{\bm{q}}_{1}}
\sum_{\bm{k}_{2}} \sum_{\bar{\bm{q}}_{2}}
\sin(k_{1z}a)\sin(k_{2z}a)  \Bigl[ N_{\rm b}^2T_{1}^2  \delta_{\bm{k}_{2\parallel},\bar{\bm{q}}_{2}} \delta_{\bm{k}_{1\parallel},\bar{\bm{q}}_{1}}
+ N_{\rm b} T_{2}^2 \delta_{{\bm k}_{1\parallel} - \bar{\bm q}_{1},{\bm k}_{2\parallel} - \bar{\bm q}_{2} }  \Bigr] \nonumber \\
& \hspace{5mm} \times
\langle \bm{k}'\gamma'; N_{\bm{k}_1}+1|b_{\bm{k}_{1}}^{\dagger}s^{x'+}_{\bar{\bm{q}}_{1}}|\bm{k}\gamma; N_{\bm{k}_1} \rangle
\langle \bm{k}\gamma; N_{\bm{k}_2}-1|
b_{\bm{k}_{2}}s^{x'-}_{\bar{\bm{q}}_{2}}|\bm{k}'\gamma';N_{\bm{k}_2}\rangle + {\rm c.c.}, \label{eq:Hint2siteend} \\
&\Bigl \langle |
\langle \bm{k}'\gamma'|
H_{\rm int,s}|\bm{k}\gamma\rangle
|^2 \Bigr\rangle_{\rm ave} \nonumber \\
&= \frac{4S_{0}^2}{N_{\rm b}^2}
\sum_{\bar{\bm{q}}_{1}} \sum_{\bar{\bm{q}}_{2}}
\Bigl[ N_{\rm b}^2 T_{1}^2  \delta_{\bar{\bm{q}}_{2},\bm{0}} \delta_{\bar{\bm{q}}_{1},\bm{0}}
+ N_{\rm b} T_{2}^2 \delta_{\bar{\bm{q}}_{1},\bar{\bm{q}}_{2}}
\Bigr]  \langle \bm{k}'\gamma'| s^{x'}_{\bar{\bm{q}}_{1}} |\bm{k}\gamma\rangle
\langle \bm{k}\gamma| (s^{x'}_{\bar{\bm{q}}_{2}})^{\dagger} |\bm{k}'\gamma'\rangle,
\label{eq:sHint2siteend}
\end{align}
where $\ket{{\bm k}\gamma; N_{\bm q}} = \ket{{\bm k} \gamma}\otimes \ket{N_{\bm q}}$ and c.c. denotes the complex conjugate.

\subsection{Momentum representation of $H_{\rm int}$}

Finally, we explain the relation of the site representation of $H_{\rm int}$ in the previous subsection with its momentum representation given in Eqs.~(\ref{eq:Hd+Hs})–(\ref{tqz}).
First, let us consider the dirty case.
The squared matrix elements are given as
\begin{align}
&|
\langle \bm{k}'\gamma' ; \{  N'_{\bm{q}}\}|
H_{\rm int,d}|\bm{k}\gamma; \{N_{\bm{q}} \} \rangle
|^2 \nonumber \\
&=\sum_{\bm{k}_{1}} \sum_{\bar{\bm{q}}_{1}}
\sum_{\bm{k}_{2}} \sum_{\bar{\bm{q}}_{2}}
2S_0 \bar{T}^2 \sin(k_{1z}a) \sin(k_{2z}a)  
\langle \bm{k}'\gamma'; N_{\bm{k}_1}+1| b^\dagger_{\bm{k}_{1}} s^{x'+}_{\bar{\bm{q}}_{1}}|\bm{k}\gamma; N_{\bm{k}_1} \rangle   \langle \bm{k}\gamma; N_{\bm{k}_2}-1| b_{\bm{k}_{2}} s^{x'-}_{\bar{\bm{q}}_{2}}|\bm{k}'\gamma'; N_{\bm{k}_2} \rangle + {\rm c.c.}, \label{eq:dtycollele2waveend}\\
&|
\langle \bm{k}'\gamma'|
H_{\rm int,s}|\bm{k}\gamma\rangle
|^2 =\sum_{\bar{\bm{q}}_{1}}
\sum_{\bar{\bm{q}}_{2}}
S_{0}^2 \bar{\mathcal{T}}^2 
\langle \bm{k}'\gamma'|s^{x'}_{\bar{\bm{q}}_{1}}|\bm{k}\gamma\rangle
\langle \bm{k}\gamma|(s^{x'}_{\bar{\bm{q}}_{2}})^{\dagger}|\bm{k}'\gamma'\rangle. \label{eq:sdtycollele2waveend}
\end{align}
By comparing these equations with with Eqs.~(\ref{eq:Hint2siteend}) and (\ref{eq:sHint2siteend}), we find that the dirty case corresponds to the condition $T_{1} \ll T_{2}$.
Then, the coefficients $\bar{T}$ and $\bar{\mathcal{T}}$ can be given as
\begin{align}
&\bar{T}^{2} = \frac{2T_{2}^{2}}{N_{\rm FI}N_{\rm b}},
\quad \bar{\mathcal{T}}^{2} = \frac{4T_{2}^{2}}{N_{\rm b}}.
\end{align}
Next, let us discuss the clean case. 
The squared matrix elements are given as
\begin{align}
|
\langle \bm{k}'\gamma';\{  N'_{\bm{q}}\} |
H_{\rm int,d}|\bm{k}\gamma; \{ N_{\bm{q}} \} \rangle
|^2 &=\sum_{\bm{k}_{1}} \sum_{\bar{\bm{q}}_{1}}
\sum_{\bm{k}_{2}} \sum_{\bar{\bm{q}}_{2}}
2S_0 \bar{T}^2 \sin(k_{1z}a) \sin(k_{2z}a) \delta_{\bm{k}_{1\parallel},\bar{\bm{q}}_{1}}
\delta_{\bm{k}_{2\parallel},\bar{\bm{q}}_{2}} \notag \\
&  \times 
\langle \bm{k}'\gamma'; N_{\bm{k}_1}+1| b^\dagger_{\bm{k}_{1}} s^{x'+}_{\bar{\bm{q}}_{1}}|\bm{k}\gamma; N_{\bm{k}_1}\rangle  \langle \bm{k}\gamma; N_{\bm{k}_1}-1| b_{\bm{k}_{2}} s^{x'-}_{\bar{\bm{q}}_{2}}|\bm{k}'\gamma'\rangle|N_{\bm{k}_2} \rangle  + {\rm h.c.}, \label{eq:clncollele2waveend}\\
|
\langle \bm{k}'\gamma'|
H_{\rm int,s}|\bm{k}\gamma\rangle
|^2 &=\sum_{\bar{\bm{q}}_{1}}
\sum_{\bar{\bm{q}}_{2}}
\bar{\mathcal{T}}^2 \delta_{\bar{\bm{q}}_{1},\bm{0}} \delta_{\bar{\bm{q}}_{2},\bm{0}}
S_{0}^2 \langle \bm{k}'\gamma'|s^{x'}_{\bar{\bm{q}}_{1}}|\bm{k}\gamma\rangle
\langle \bm{k}\gamma|(s^{x'}_{\bar{\bm{q}}_{2}})^{\dagger}|\bm{k}'\gamma'\rangle. \label{eq:sclncollele2waveend}
\end{align}
By comparing these equations with Eqs.~(\ref{eq:Hint2siteend}) and (\ref{eq:sHint2siteend}), we find that the clean case corresponds to the condition $T_1 \gg T_2$.
Then, the coefficients, $\bar{T}$ and $\bar{\cal T}$, are given as
\begin{align}
&\bar{T}^{2} = \frac{2T_{1}^{2}}{N_{\rm FI}},
\quad \bar{\mathcal{T}}^{2} = 4T_{1}^{2}.
\end{align}
Thus, the present simple modeling can relate the coupling constant with the energy of the interfacial exchange coupling, directly.

\section{Derivation of collision terms}
\label{app:CollisionTerms}

In this appendix, we show explicit forms of the collision terms which appear in the Boltzmann equation.
First, we show the collision term due to the nonmagnetic impurities:
\begin{align}
\left. \frac{\partial f(\bm{k},\gamma)}{\partial t} \right|_{\rm imp }
=\sum_{\bm{k}'}\sum_{\gamma'=\pm}
\Bigl{[}
P_{\bm{k}'\gamma'\rightarrow\bm{k}\gamma}
f(\bm{k}',\gamma')(1-f(\bm{k},\gamma))-P_{\bm{k}\gamma\rightarrow\bm{k}'\gamma'}
f(\bm{k},\gamma)(1-f(\bm{k}',\gamma'))
\Bigl{]},
\end{align}
where $P_{\bm{k}\gamma\rightarrow\bm{k}'\gamma'}$ represents the electron transition rate due to impurity scattering and can be written using the Born approximation as follows:
\begin{align}
&P_{\bm{k}\gamma\rightarrow\bm{k}'\gamma'}
= \frac{2\pi}{\hbar}
|\langle \bm{k}'\gamma'|H_{\rm imp}(\{{\bm R}_i\})|\bm{k}\gamma\rangle |^2 \delta(E^{\gamma'}_{\bm{k}'}-E^{\gamma}_{\bm{k}}), \label{eq:Qkkp}
\end{align}
where $H_{\rm imp}(\{{\bm R}_i\})$ is a scattering matrix, whose matrix element is given as
$\braket{{\bm k}'\sigma'| H_{\rm imp}(\{{\bm R}_i\}) |{\bm k}\sigma} = (u/{{\cal A}}) \delta_{\sigma,\sigma'}\sum_ie^{-i({\bm k}'-{\bm k})
\cdot {\bm R}_i}$ ($\sigma,\sigma' = \uparrow, \downarrow$).
After averaging over the impurity positions $\{ {\bm R}_i \}$, the collision term is calculated as 
\begin{align} 
\left. 
\frac{\partial f(\bm{k},\gamma)}{\partial t}\right|_{\rm imp}
=\frac{2\pi u^2 n_{\rm imp}}{\hbar\mathcal{A}}\sum_{\bm{k}',\gamma'}\sum_{\sigma,\sigma'}
C^*_{\sigma\gamma'}(\varphi')C_{\sigma\gamma}(\varphi)C_{\sigma'\gamma'}(\varphi')C^*_{\sigma'\gamma}(\varphi)
[f(\bm{k}',\gamma')-f(\bm{k},\gamma)]\delta(E^{\gamma'}_{\bm{k}'}-E^{\gamma}_{\bm{k}}), \label{eq:impCC}
\end{align}
where $n_{\rm imp}$ is an impurity density per unit area.
Using $C_{\uparrow \gamma}(\varphi)= C(\varphi)/\sqrt{2}$ and $C_{\downarrow \gamma}(\varphi) = \gamma/\sqrt{2}$ with Eq.~(\ref{eq:ckstg})
the collision term is calculated as
\begin{align}
\left. \frac{\partial f(\bm{k},\gamma)}{\partial t}\right|_{\rm imp}
=\frac{\pi u^2 n_{\rm imp}}{\hbar\mathcal{A}}
\sum_{\bm{k}',\gamma'}[1+\gamma\gamma'\hat{\bm{h}}_{\rm eff}(\varphi)\cdot\hat{\bm{h}}_{\rm eff}(\varphi')]
[f(\bm{k}',\gamma')-f(\bm{k},\gamma)]\delta(E^{\gamma'}_{\bm{k}'}-E^{\gamma}_{\bm{k}}), \label{eq:appgoldenimp}
\end{align}
where $\hat{\bm{h}}_{\rm eff}(\varphi)=\bm{h}_{\rm eff}(\bm{k})/|\bm{h}_{\rm eff}(\bm{k})|$ represents the direction of the effective Zeeman field generated by the Rashba and Dresselhaus spin-orbit interactions.
Using the formula of Eq.~(\ref{eq:sumint}), the summation with respect to ${\bm k}'$ can be replace with an integral as
\begin{align}
\left. \frac{\partial f(\bm{k},\gamma)}{\partial t} \right|_{\rm imp} 
=\frac{\Gamma}{4\pi\hbar D(\epsilon_{\rm F})}\int_{0}^{\infty}\! dk'\, |{\bm k}'|\int_{0}^{2\pi}\! \!\frac{d\varphi'}{2\pi}
\sum_{\gamma'}[1+\gamma\gamma'\hat{\bm{h}}_{\rm eff}(\varphi)\cdot\hat{\bm{h}}_{\rm eff}(\varphi')]
[f(\bm{k}',\gamma')-f(\bm{k},\gamma)] \delta(E^{\gamma'}_{\bm{k}'}-E^{\gamma}_{\bm{k}}),\label{eq:Eximpbol} 
\end{align}
where $\Gamma=2\pi n_{\rm imp}u^{2}D(\epsilon_{\rm F})$.

In the same way, the collision term due to the interfacial exchange coupling can be constructed as
\begin{align}
\left. \frac{\partial f(\bm{k},\gamma)}{\partial t} \right|_{\rm int} =\sum_{\bm{k}'}\sum_{\gamma'}
\Bigl{[}
Q_{\bm{k}'\gamma'\rightarrow\bm{k}\gamma}
f(\bm{k}',\gamma')(1-f(\bm{k},\gamma))-Q_{\bm{k}\gamma\rightarrow\bm{k}'\gamma'}
f(\bm{k},\gamma)(1-f(\bm{k}',\gamma'))
\Bigl{]},\label{eq:fimpintborn}
\end{align}
where $Q_{\bm{k}\gamma\rightarrow\bm{k}'\gamma'}$ denotes the transition rate due to the interfacial exchange coupling, which is written as
\begin{align}
Q_{\bm{k},\gamma\rightarrow \bm{k}',\gamma'}&=
\sum_{\bm{q},\bm{q}'}
\sum_{N_{\bm{q}},N'_{\bm{q}'}}
\frac{2\pi}{\hbar}{|}
\langle \bm{k}'\gamma'|\langle N'_{\bm{q}'}|
H_{\rm int,d}|\bm{k}\gamma\rangle|N_{\bm{q}} \rangle
{|}^{2}\delta(E^{\gamma'}_{\bm{k}'}+N'_{\bm{q}'}\hbar\omega_{\bm{q}'}-E^{\gamma}_{\bm{k}}-N_{\bm{q}}\hbar\omega_{\bm{q}})
\rho(N_{\bm{q}})\nonumber \\
&\hspace{5mm}+\frac{2\pi}{\hbar}|\langle \bm{k}'\gamma'|H_{\rm int,s}|\bm{k}\gamma\rangle|^2 \delta(E^{\gamma'}_{\bm{k}'}-E^{\gamma}_{\bm{k}}).
\label{eq:0Pkgam}
\end{align}
Here, $H_{\rm int,d}$ and $H_{\rm int,s}$ are matrices describing the interfacial scattering,
$|N_{\bm{q}}\rangle$ represents the eigenstate of the magnon number operator, and $\rho(N_{\bm{q}})$ is given by $\rho(N_{\bm{q}})=e^{-\upbeta\hbar\omega_{\bm{q}}N_{\bm{q}}}/\sum_{N_{\bm{q}}=0}^{\infty}e^{-\upbeta\hbar\omega_{\bm{q}}N_{\bm{q}}}$.
For further calculations, we need to fix the matrix elements of $H_{\rm int}$ depending on the kind of interface as shown in Eqs.~(\ref{eq:dirty}) and (\ref{eq:clean}).

\section{Derivation of Eq.~(\ref{eq:fsolimp1})}
\label{app1}

In this appendix, we derive the result of the distribution function for the direct Rashba-Edelstein effect, Eq.~(\ref{eq:fsolimp1}).
Substituting the expression of the collision term due to impurity scattering given in Appendix~\ref{app:CollisionTerms} into Eq.~(\ref{eq:REequation}), we obtain the following integral equation with respect to $f_1(\bm{k},\gamma)$:
\begin{align}
f_1(\bm{k},\gamma)
&=\mathcal{L}(\bm{k},\gamma)+\frac{\hbar^{2}}{2m^{*}}\int_{0}^{2\pi}\frac{d\varphi'}{2\pi}
\sum_{\gamma'}\int_{0}^{\infty}dk'\, |\bm{k}'|[1+\gamma\gamma'\hat{\bm{h}}_{\rm eff}(\varphi)\cdot\hat{\bm{h}}_{\rm eff}(\varphi')]f_1(\bm{k}',\gamma')\delta(E^{\gamma'}_{\bm{k}'}-E^{\gamma}_{\bm{k}}),\\
\mathcal{L}(\bm{k},\gamma)
&=-\frac{\hbar eE_{x}}{\Gamma}\cdot \frac{\partial f_{0}(E^{\gamma}_{\bm k})}{\partial E^{\gamma}_{\bm{k}}}\Bigl{(}
\frac{\hbar |\bm{k}|}{m^{*}}\cos\varphi
+\frac{\gamma}{\hbar\sqrt{\alpha^2+\beta^2+2\alpha \beta \sin 2\varphi}}[(\alpha^{2}+\beta^{2})\cos\varphi+2\alpha\beta\sin\varphi]
\Bigl{)}
\end{align}
Successive substitution of $f_1({\bm k},\gamma)$ into the right-hand side yields
\begin{align}
f_1(\bm{k},\gamma)
&=\mathcal{L}(\bm{k},\gamma)
+\frac{\hbar^{2}}{2m^{*}}
\int_{0}^{2\pi}\!\frac{d\varphi''}{2\pi}
\sum_{\gamma''} \gamma\gamma'' \int_{0}^{\infty}\! dk''\, |\bm{k}''|\,
\delta(E^{\gamma''}_{\bm{k}''}-E^{\gamma}_{\bm{k}}) \nonumber \\
&\hspace{10mm} \times 
\hat{\bm{h}}_{\rm eff}^{T}(\varphi)
\biggl{(}\hat{I}-\int_{0}^{2\pi}\frac{d\varphi'}{2\pi}\hat{\bm{h}}_{\rm eff}(\varphi')\cdot \hat{\bm{h}}_{\rm eff}^{T}(\varphi')\biggl{)}^{-1}
\hat{\bm{h}}_{\rm eff}(\varphi'')
\mathcal{L}(\bm{k}'',\gamma'').\label{eq:f1LML}
\end{align}
Here, $\hat{I}$ represents the identity matrix, $\hat{A}^{-1}$ denotes the inverse matrix of $\hat{A}$, and $\bm{a} \cdot \bm{a}^T$ is defined by the following matrix expression:
\begin{align}
\bm{a} \cdot \bm{a}^T = \left( \begin{array}{c} a_x \\ a_y \end{array} \right) (a_x \ a_y) 
= \left( \begin{array}{cc} a_x a_x & a_x a_y \\ a_y a_x & a_y a_y \end{array} \right) .
\end{align}
The second term of the right-hand side in Eq.~(\ref{eq:f1LML}) can be calculated as
\begin{align}
&\frac{\hbar^{2}}{2m^{*}}
\int_{0}^{2\pi}\!\frac{d\varphi''}{2\pi}
\sum_{\gamma''} \gamma\gamma'' \int_{0}^{\infty}\! dk''\, |\bm{k}''|\,
\delta(E^{\gamma''}_{\bm{k}''}-E^{\gamma}_{\bm{k}})
\hat{\bm{h}}_{\rm eff}^{T}(\varphi)
\biggl{(}\hat{I}-\int_{0}^{2\pi}\frac{d\varphi'}{2\pi}\hat{\bm{h}}_{\rm eff}(\varphi')\cdot \hat{\bm{h}}_{\rm eff}^{T}(\varphi')\biggl{)}^{-1}
\hat{\bm{h}}_{\rm eff}(\varphi'')
\mathcal{L}(\bm{k}'',\gamma'')\nonumber \\
&=\frac{eE_{x}}{\Gamma}\cdot \frac{\partial f_{0}(E^{\gamma}_{\bm k})}{\partial E^{\gamma}_{\bm{k}}}
\cdot\frac{\gamma}{\sqrt{\alpha^2+\beta^2+2\alpha \beta \sin 2\varphi}}[(\alpha^{2}+\beta^{2})\cos\varphi+2\alpha\beta\sin\varphi]. \label{eq:stermsim}
\end{align}
Here, we used
\begin{align}
&\delta(E^{\gamma'}_{\bm{k}'}-E^{\gamma}_{\bm{k}})
\simeq \frac{m^{*}}{\hbar^{2}
\sqrt{2m^* E^{\gamma}_{\bm{k}}/\hbar^2}}
\delta(k'-k'(\bm{k},\gamma,\varphi',\gamma')),\label{eq:delE} \\
&k'(\bm{k},\gamma,\varphi',\gamma')
\simeq\sqrt{2m^{*}E^{\gamma}_{\bm{k}}/\hbar^2}
-\frac{m^{*}\gamma'\sqrt{\alpha^2+\beta^2+2\alpha\beta \sin 2\varphi'}}{\hbar^{2}},\label{eq:defkp}
\end{align}
where second-order terms of the spin-orbit interaction were dropped. Substituting Eq.~(\ref{eq:stermsim}) into Eq.~(\ref{eq:f1LML}) yields the following solution:
\begin{align}
f_1(\bm{k},\gamma)=-\frac{\partial f_{0}(E^{\gamma}_{\bm k})}{\partial E^{\gamma}_{\bm{k}}}\frac{\hbar^2 e E_x |\bm{k}|}{\Gamma m^*}\cos\varphi. \label{eq:f1sol}
\end{align}
We note that this solution satisfies the charge conservation condition, $\sum_{{\bm k},\gamma} f_1({\bm k},\gamma) = 0$. Comparing Eq.~(\ref{eq:f1sol}) with Eq.~(\ref{eq:f1org}) allows us to get
\begin{align}
\delta \mu_1(\bm{k},\gamma)= \frac{\hbar^2 e E_x |\bm{k}|}{\Gamma m^*}\cos\varphi.\label{eq:mu1ends}
\end{align}
Thus, Eq.~(\ref{eq:fsolimp1}) can be derived.

\section{Detailed calculation for the dirty interface}
\label{app:dirty}

For the dirty interface, the scattering matrix is expressed as $H_{\rm int} = H_{\rm int,s} + H_{\rm int,d}$, where $H_{\rm int,s}$ describes a static contribution due to the exchange bias and $H_{\rm int,d}$ describes a dynamic contribution accompanying magnon absorption or emission.
These matrices, which act on both of the Hilbert spaces for conduction electrons and magnons, are given as
\begin{align}
\braket{{\bm k}'\sigma'| H_{\rm int,s} |{\bm k}\sigma} &= \frac{S_0 \bar{\cal T}}{2} 
(\hat{\sigma}^{x'})_{\sigma'\sigma}
, \\
\braket{{\bm k}'\sigma'| H_{\rm int,d} |{\bm k}\sigma} &= \frac{\sqrt{2S_0}\bar{T}}{2} (\hat{\sigma}^{x'-})_{\sigma' \sigma}\sum_{\bm q} \sin(q_z a)b_{{\bm q}} + \frac{\sqrt{2S_0}\bar{T}^*}{2} (\hat{\sigma}^{x'+})_{\sigma' \sigma}\sum_{\bm q} \sin(q_z a) b_{{\bm q}}^\dagger,
\end{align}
and $b_{{\bm q}}^\dagger$ and $b_{{\bm q}}$ are creation and annihilation operators of magnons.
By the basis transformation, the matrix elements are rewritten with an energy eigenbasis $\ket{{\bm k}\gamma}$ as
\begin{align}
\braket{{\bm k}'\gamma'| H_{\rm int,s} |{\bm k}\gamma}
&= \frac{S_0 \bar{\cal T}}{2} \sum_{\sigma,\sigma'} C_{\sigma'\gamma'}^*({\bm k}')
(\hat{\sigma}^{x'})_{\sigma'\sigma}
C_{\sigma\gamma}({\bm k})  , \\
\braket{{\bm k}'\gamma'| H_{\rm int,d} |{\bm k}\gamma} &= \frac{\sqrt{2S_0}\bar{T}}{2} \sum_{\sigma,\sigma'}C_{\sigma'\gamma'}^*({\bm k}') (\hat{\sigma}^{x'-})_{\sigma' \sigma} C_{\sigma\gamma}({\bm k}) \sum_{\bm q}\sin(q_z a) b_{{\bm q}} \nonumber \\
&\hspace{5mm}+ \frac{\sqrt{2S_0}\bar{T}^*}{2}\sum_{\sigma,\sigma'} C_{\sigma'\gamma'}^*({\bm k}') (\hat{\sigma}^{x'+})_{\sigma' \sigma} C_{\sigma\gamma}({\bm k}) \sum_{\bm q}\sin(q_z a) b_{{\bm q}}^\dagger,
\end{align}
Substituting these matrix elements into the collision term due to the interfacial scattering given in Appendix~\ref{app:CollisionTerms}, we obtain
\begin{align}
\left. \frac{\partial f(\bm{k},\gamma)}{\partial t} \right|_{\rm int}
&=-\frac{\pi S_{0}|\bar{T}|^{2}}{\hbar}
\sum_{\bm{k}',\gamma'}\sum_{\bm{q}}
\langle N_{\bm{q}}\rangle\sin^2(q_z a)
\Bigl{(}[1-\gamma\hat{\bm{h}}_{\rm eff}(\varphi)\cdot\hat{\bm{m}}][1+\gamma'\hat{\bm{h}}_{\rm eff}(\varphi')\cdot\hat{\bm{m}}]
A(\bm{k},\gamma,\bm{k}',\gamma')\delta(E^{\gamma'}_{\bm{k}'}-E^{\gamma}_{\bm{k}}-\hbar\omega_{\bm{q}})
\nonumber \\
&\hspace{30mm}-[1+\gamma\hat{\bm{h}}_{\rm eff}(\varphi)\cdot\hat{\bm{m}}][1-\gamma'\hat{\bm{h}}_{\rm eff}(\varphi')\cdot\hat{\bm{m}}]
A(\bm{k}',\gamma',\bm{k},\gamma)\delta(E^{\gamma'}_{\bm{k}'}-E^{\gamma}_{\bm{k}}
+\hbar\omega_{\bm{q}})
\Bigl{)}\nonumber \\
&\hspace{-10mm}
-\frac{\pi S_{0}^{2}|\bar{\mathcal{T}}|^{2}}{4\hbar}
\sum_{\bm{k}',\gamma'}
\Bigl{(}
1+
2\gamma\gamma'
[\hat{\bm{h}}_{\rm eff}(\varphi)\cdot\hat{\bm{m}}]
[\hat{\bm{h}}_{\rm eff}(\varphi')\cdot\hat{\bm{m}}]
-\gamma\gamma'\hat{\bm{h}}_{\rm eff}(\varphi)\cdot\hat{\bm{h}}_{\rm eff}(\varphi')
\Bigl{)}A(\bm{k},\gamma,\bm{k}',\gamma')\delta(E^{\gamma'}_{\bm{k}'}-E^{\gamma}_{\bm{k}}),\label{eq:pfimpintsurmag}
\end{align}
where $A(\bm{k},\gamma,\bm{k}',\gamma')=
\upbeta f_{0}(\bm{k},\gamma)[1-f_{0}(\bm{k}',\gamma')]
[\delta\mu(\bm{k},\gamma)-\delta\mu(\bm{k}',\gamma')]$,
$\hat{\bm{m}}=(\cos\theta, \sin\theta)^T$ represents the direction of the spin polarization in the FI, $N_{\rm FI}$ denotes the number of unit cells within the FI, and $\delta\mu(\bm{k}, \gamma)$ is the chemical potential shift.
Hereafter, we assume that $\hbar\omega_{\bm{q}}$ is small and approximated $\delta(E^{\gamma'}_{\bm{k}'}-E^{\gamma}_{\bm{k}}
\pm\hbar\omega_{\bm{q}})$ as $\delta(E^{\gamma'}_{\bm{k}'}-E^{\gamma}_{\bm{k}})$.

Next, assuming that the interfacial scattering is sufficiently weak, we use the perturbative relation, Eq.~(\ref{eq:OET2}).
Then, we obtain the following integral equation for $f_{\rm D}(\bm{k},\gamma)$:
\begin{align}
f_{\rm D}(\bm{k},\gamma) &=\mathcal{G}_{\rm D}(\bm{k},\gamma,\theta)+\frac{\hbar^{2}}{2m^{*}}\int_{0}^{2\pi}\!\frac{d\varphi'}{2\pi}
\sum_{\gamma'=\pm}\int_{0}^{\infty}\!dk'\,|\bm{k}'|
[1+\gamma\gamma'\hat{\bm{h}}_{\rm eff}(\varphi)\cdot\hat{\bm{h}}_{\rm eff}(\varphi')]
 \delta(E^{\gamma'}_{\bm{k}'}-E^{\gamma}_{\bm{k}}) f_{\rm D}(\bm{k}',\gamma'),\label{eq:fmagbolint} \\
\mathcal{G}_{\rm D}(\bm{k},\gamma,\theta)&=\frac{\pi D(\epsilon_{\rm F})S_{0}\hbar^{2} eE_{x}\mathcal{A}}{2\Gamma^{2}m^{*}}
\frac{\partial f_{0}(E^{\gamma}_{\bm{k}})}{\partial E^{\gamma}_{\bm{k}}}
\int_{0}^{2\pi}\frac{d\varphi'}{2\pi}\nonumber \\
&\times \Bigl{[}
8|\bar{T}|^{2}\sum_{\bm{q}}\langle N_{\bm{q}} \rangle \sin^2(q_z a)
\Bigl{(}
|\bm{k}| \cos\varphi-\gamma[\hat{\bm{h}}_{\rm eff}(\varphi)\cdot\hat{\bm{m}}(\theta)][\hat{\bm{h}}_{\rm eff}(\varphi')\cdot\hat{\bm{m}}(\theta)]
\frac{2m^{*}\kappa(\varphi')}{\hbar^{2}}\cos\varphi'
\Bigl{)}\nonumber \\
&\hspace{0mm}
+S_{0}|\bar{\mathcal{T}}|^{2}
\Bigl{(}
|\bm{k}| \cos\varphi
+\gamma
\{2[\hat{\bm{h}}_{\rm eff}(\varphi)\cdot\hat{\bm{m}}(\theta)]
[\hat{\bm{h}}_{\rm eff}(\varphi')\cdot\hat{\bm{m}}(\theta)]-\hat{\bm{h}}_{\rm eff}(\varphi)\cdot\hat{\bm{h}}_{\rm eff}(\varphi')\}
\frac{2m^{*}\kappa(\varphi')}{\hbar^{2}}\cos\varphi'
\Bigl{)}\Bigl{]}
\label{eq:def:mathG},
\end{align}
where $\kappa(\varphi)=\sqrt{\alpha^2+\beta^2+2\alpha\beta\sin2\varphi}$.
Iterative substitution of $f_{\rm D}({\bm k},\gamma)$ into the right-hand side of Eq.~(\ref{eq:fmagbolint}) yields
\begin{align}
f_{\rm D}(\bm{k},\gamma)&=\mathcal{G}_{\rm D}(\bm{k},\gamma,\theta)
+\frac{\hbar^{2}}{2m^{*}}
\int_{0}^{2\pi}\!\frac{d\varphi''}{2\pi}
\sum_{\gamma''} \gamma\gamma'' \int_{0}^{\infty}\! dk''\, |\bm{k}''|\,
\delta(E^{\gamma''}_{\bm{k}''}-E^{\gamma}_{\bm{k}}) \nonumber \\
&\hspace{10mm} \times 
\hat{\bm{h}}_{\rm eff}^{T}(\varphi)
\biggl{(}\hat{I}-\int_{0}^{2\pi}\frac{d\varphi'}{2\pi}\hat{\bm{h}}_{\rm eff}(\varphi')\cdot \hat{\bm{h}}_{\rm eff}^{T}(\varphi')\biggl{)}^{-1}
\hat{\bm{h}}_{\rm eff}(\varphi'')
\mathcal{G}_{\rm D}(\bm{k}'',\gamma'',\theta).\label{eq:appGfmag}
\end{align}
Here, $\hat{I}$ and $\hat{A}^{-1}$ represent the identity matrix and the inverse matrix of $\hat{A}$, respectively.
By specifically calculating Eq.~(\ref{eq:appGfmag}) and retaining only the parts dependent on $\theta$, we obtain the following:
\begin{align}
f_{\rm D}(\bm{k},\gamma)&=
\gamma\frac{\partial f_{0}(E^{\gamma}_{\bm{k}})}{\partial E^{\gamma}_{\bm{k}}}
\hat{\bm{h}}_{\rm eff}(\varphi)
\cdot\bm{V}(\theta),\label{eq:ftinthV}\\
\bm{V}(\theta)
&=
\frac{\pi D(\epsilon_{\rm F})S_{0} eE_{x}\mathcal{A}}{\Gamma^{2}}
\Bigl{[}
-8|\bar{T}|^{2}
\sum_{\bm{q}}\langle N_{\bm{q}}\rangle\sin^2(q_z a)
+2S_{0}|\mathcal{\bar{T}}|^{2}
\Bigl{]}\nonumber \\
&\times
\int_{0}^{2\pi}\frac{d\varphi''}{2\pi}
[\hat{\bm{h}}_{\rm eff}(\varphi'')\cdot\hat{\bm{m}}]
\kappa(\varphi'')\cos\varphi''
\biggl{(}\hat{I}-\int_{0}^{2\pi}\frac{d\varphi'}{2\pi}\hat{\bm{h}}_{\rm eff}(\varphi')\cdot \hat{\bm{h}}_{\rm eff}^{T}(\varphi')\biggl{)}^{-1}
\hat{\bm{m}}.\label{def:vecV}
\end{align}
Using the direct calculation of the matrix 
\begin{align}
\hat{M}\equiv \biggl{(}\hat{I}-\int_{0}^{2\pi}\frac{d\varphi'}{2\pi}\hat{\bm{h}}_{\rm eff}(\varphi')\cdot \hat{\bm{h}}_{\rm eff}^{T}(\varphi')\biggl{)}^{-1} = \frac{2}{1-\eta^2} \left( \begin{array}{cc} 1 & -\eta \\ -\eta & 1 \end{array} \right),
\end{align}
the analytical solution given in Eqs.~(\ref{eq:mu2result})-(\ref{def:eta}) can be derived.
To clarify the modulation of the distribution function, the chemical potential shift $\delta\mu_{\rm D}(\varphi,\gamma=+)$ is plotted in Fig.~\ref{fig:delmu+D}, where we omitted the term proportional to $|\bar{T}|^2$ as in Eq.~(\ref{eq: defITSNT}) and used the following constant for normalization:
\begin{align}
\mu_{x,{\rm D}}=  -\frac{2\pi D(\epsilon_{\rm F})S_{0}^{2} eE_{x}\mathcal{A}|\bar{\mathcal{T}}|^{2} x}{\Gamma^{2}}, \quad (x=\alpha,\beta).
\end{align}
These plots are consistent with the schematic diagrams for the modulation of the distribution functions shown in 
Fig.~\ref{fig:Rashba}(c) and Fig.~\ref{fig:aob11}(c).

Using the definition given in Eq.~(\ref{eq:defs}), the modulation of the spin density in the 2DEG is expressed with $f_{\rm D}({\bm k},\gamma)$ as
\begin{align}
\Delta {\bm s}_{\rm D} = \frac{\hbar}{2{\cal A}} \sum_{{\bm k},\gamma} \braket{{\bm k}\gamma|\hat{\bm{\sigma}}|{\bm k}\gamma} f_{\rm D}({\bm k},\gamma).
\end{align}
By using the analytic solution for $f_{\rm D}({\bm k},\gamma)$ and by replacing the summation with an integral, the spin density modulation is calculated as
\begin{align}
\Delta\bm{s}_{\rm D}
&=\frac{k_{\rm F}}{2\pi v_{\rm F}}
\int_0^{2\pi} \frac{d\varphi}{2\pi}
\hat{\bm{h}}_{\rm eff}(\varphi)
[\hat{\bm{h}}_{\rm eff}(\varphi)
\cdot\bm{V}(\theta)] \nonumber \\
&= \frac{k_{\rm F}D(\epsilon_{\rm F})S_{0} eE_{x}\mathcal{A}}{2v_{\rm F}\Gamma^{2}}
\Bigl{[}
-4|\bar{T}|^{2}
\sum_{\bm{q}}\langle N_{\bm{q}}\rangle \sin^2(q_z a)
+S_{0}|\mathcal{\bar{T}}|^{2}
\Bigl{]}
\frac{\alpha\sin\theta-\beta\cos\theta}{1-\eta^{2}}
\begin{pmatrix}
\cos\theta & \sin\theta \\
\sin\theta & \cos\theta
\end{pmatrix}
\begin{pmatrix}
1+\eta^{2} \\ -2\eta
\end{pmatrix},\label{eq:sREMRendA}
\end{align}
where we used the following approximation:
\begin{align}
\frac{\partial f_{0}(E^{\gamma}_{\bm{k}})}{\partial E^{\gamma}_{\bm{k}}}
\simeq-\delta(E^{\gamma}_{\bm{k}}-\mu).\label{eq:pfpEdel}
\end{align}
Thus, Eq.~(\ref{eq:sREMRend}) can be derived.
In a similar way, using the definition given in Eqs.~(\ref{jREMR}) and (\ref{eq:defv}), the modulation of the current density is calculated as
\begin{align}
\Delta\bm{j}_{\rm D}
&=\frac{e}{\mathcal{A}}\sum_{\gamma=\pm}\sum_{\bm{k}} \bm{v}(\bm{k},\gamma) f_{\rm D}(\bm{k},\gamma) \nonumber \\
&=\frac{ek_{\rm F}}{\pi\hbar^{2} v_{\rm F}}
\int_{0}^{2\pi}\frac{d\varphi}{2\pi}\frac{\hat{\bm{h}}_{\rm eff}(\varphi)\cdot\bm{V}(\theta)}{\kappa(\varphi)}
\left( \begin{array}{c} 
(\alpha^{2}+\beta^{2})\cos\varphi
+2\alpha\beta\sin3\varphi \\(\alpha^{2}+\beta^{2})\sin\varphi-2\alpha\beta\cos3\varphi \end{array} \right).\label{eq:jREMRfinhV}
\end{align}
Here, Eq.~(\ref{eq:pfpEdel}) was also used. Substituting Eq.~(\ref{def:vecV}), the current density modulation is rewritten as
\begin{align}
\Delta\bm{j}_{\rm D}
=\frac{e^{2}k_{\rm F}D(\epsilon_{\rm F})S_{0}E_{x}\mathcal{A}}{\hbar^{2} v_{\rm F}\Gamma^{2}}
\Bigl{[}
-4|\bar{T}|^{2}
\sum_{\bm{q}}\langle N_{\bm{q}}\rangle\sin^2(q_z a)
+S_{0}|\mathcal{\bar{T}}|^{2}
\Bigl{]}
\left( \begin{array}{c}
(\alpha\sin\theta-\beta\cos\theta)^{2} \\
-(\alpha^{2}+\beta^{2})\cos\theta\sin\theta+\alpha\beta
\end{array} \right)
.\label{eq:REMRendxy}
\end{align}
By subtracting terms independent of $\theta$, Eq.~(\ref{eq:REMRendxynew}) can be derived.

\begin{figure*}[tb]
\centering
\includegraphics[width=145mm]{Fig7.pdf}
\caption{Left panels: The chemical potential shift
$\delta\mu_{\rm D}(\varphi,+)-\delta\mu_{\rm D}(\varphi_0,+)$ for the dirty interface is plotted as a function of the azimuth angle $\varphi$ of the electron wavenumber for $\beta = 0$ (top plot) and $\alpha/\beta = 1.1$ (bottom plot).
The reference point of the chemical potential shift is taken as $\varphi_0=\pi/2,3\pi/2$ for the former case, while as $\varphi_0=3\pi/4,7\pi/4$ for the latter case.
Note that $\gamma=+$ corresponds to the inner Fermi surfaces.
Right panels: Schematic diagrams for modulation of the distribution functions for $\beta = 0$ (the upper diagrams) and $\alpha/\beta = 1.1$ (the lower diagrams).
The orange (blue) regions represent the places where the distribution function increases (decreases) compared to the reference point.
}
\label{fig:delmu+D}
\end{figure*}

\section{Detailed calculation for the clean interface}
\label{app:clean}

For the clean interface, the scattering matrix is expressed as $H_{\rm int} = H_{\rm int,s} + H_{\rm int,d}$, where
\begin{align}
\braket{{\bm k}'\sigma'| H_{\rm int,s} |{\bm k}\sigma} &= \frac{S_0 \bar{\cal T}}{2} 
(\hat{\sigma}^{x'})_{\sigma'\sigma}
\delta_{{\bm k},{\bm k}'} \\
\braket{{\bm k}'\sigma'| H_{\rm int,d} |{\bm k}\sigma} &= \frac{\sqrt{2S_0}\bar{T}}{2} (\hat{\sigma}^{x'-})_{\sigma' \sigma} \sum_{\bm q} b_{{\bm q}}\sin(q_z a)\delta_{{\bm q}_\parallel,{\bm k}'-{\bm k}} + \frac{\sqrt{2S_0}\bar{T}^*}{2} (\hat{\sigma}^{x'+})_{\sigma' \sigma}\sum_{\bm q} b_{{\bm q}}^\dagger\sin(q_z a)\delta_{{\bm q}_\parallel,{\bm k}-{\bm k}'} ,
\end{align}
and $b_{{\bm q}}^\dagger$ and $b_{{\bm q}}$ are creation and annihilation operators of magnons.
By the basis transformation, the matrix elements are rewritten with an energy eigenbasis $\ket{{\bm k}\gamma}$ as
\begin{align}
\braket{{\bm k}'\gamma'| H_{\rm int,s} |{\bm k}\gamma} 
&= \frac{S_0 \bar{\cal T}}{2} \sum_{\sigma,\sigma'} C_{\sigma'\gamma'}^*({\bm k}')
(\hat{\sigma}^{x'})_{\sigma'\sigma}
C_{\sigma\gamma}({\bm k}) \delta_{{\bm k},{\bm k}'}  , \\
\braket{{\bm k}'\gamma'| H_{\rm int,d} |{\bm k}\gamma} &= \frac{\sqrt{2S_0}\bar{T}}{2} \sum_{\sigma,\sigma'}C_{\sigma'\gamma'}^*({\bm k}') (\hat{\sigma}^{x'-})_{\sigma' \sigma} C_{\sigma\gamma}({\bm k}) \sum_{\bm q} b_{{\bm q}}\sin(q_z a)\delta_{{\bm q}_\parallel,{\bm k}'-{\bm k}} \nonumber \\
&\hspace{5mm}+\frac{\sqrt{2S_0}\bar{T}^*}{2} \sum_{\sigma,\sigma'}C_{\sigma'\gamma'}^*({\bm k}') (\hat{\sigma}^{x'+})_{\sigma' \sigma} C_{\sigma\gamma}({\bm k}) \sum_{\bm q} b_{{\bm q}}^{\dagger}\sin(q_z a)\delta_{{\bm q}_\parallel,{\bm k}-{\bm k}'}.
\end{align}
We note that, compared with the dirty case, the Kronecker delta in the last part of each term is added because of the in-plane momentum conservation law.

The subsequent calculation is almost the same as that for the dirty interface, except for the in-plane momentum conservation law.
The Boltzmann equation can be written in the form of the integral equation as
\begin{align}
f_{\rm C}(\bm{k},\gamma) &= \mathcal{G}_{\rm C}(\bm{k},\gamma,\theta) +\frac{\hbar^{2}}{2m^{*}}\int_{0}^{2\pi}\frac{d\varphi'}{2\pi}
\sum_{\gamma'}\int_{0}^{\infty}dk'\,|\bm{k}'|
[1+\gamma\gamma'\hat{\bm{h}}_{\rm eff}(\varphi)\cdot\hat{\bm{h}}_{\rm eff}(\varphi')]\delta(E^{\gamma'}_{\bm{k}'}-E^{\gamma}_{\bm{k}})
f_{\rm C}(\bm{k}',\gamma'),\label{eq:fmagbolintmain} \\
\mathcal{G}_{\rm C}(\bm{k},\gamma,\theta)&\simeq \frac{4\hbar^{2}\pi D(\epsilon_{\rm F})S_0 |\bar{T}|^{2}\mathcal{A}eE_x}{\Gamma^2 m^{*}} 
\frac{\partial f_{0}(E^{\gamma}_{\bm{k}})}{\partial E^{\gamma}_{\bm{k}}}
\sum_{q_{z}>0}
\int_{0}^{2\pi}\frac{d\varphi'}{2\pi}
N(\varphi-\varphi',q_{z})\sin^2(q_z a)
\mathcal{F}({\bm k},\gamma,\varphi',\theta), \\
\mathcal{F}({\bm k},\gamma,\varphi',\theta) &=|\bm{k}|\cos\varphi
-\sqrt{2m^{*}E^{\gamma}_{\bm{k}}/\hbar^{2}}\cos\varphi'\nonumber \\
& +\frac{m^{*}h_{\rm eff}(\varphi')}{\hbar^{2}k_{\rm F}}\gamma[\hat{\bm{h}}_{\rm eff}(\varphi)\cdot\hat{\bm{m}}(\theta)][\hat{\bm{h}}_{\rm eff}(\varphi')\cdot\hat{\bm{m}}(\theta)]\Bigl{(}\frac{|\bm{k}|\cos\varphi}{\sqrt{2m^{*}E^{\gamma}_{\bm{k}}/\hbar^{2}}}-2\cos\varphi'\Bigl{)},\label{Gapp4}
\end{align}
where $N(\varphi,q_{z})$ is the magnon distribution function, which is given in Eq.~(\ref{eq:hoNdef}).
Here, assuming that the magnon energy $\hbar\omega_{\bm{q}} = \hbar \omega_{{\bm q}_{\parallel},q_z}$ is much smaller than the spin-splitting energy, we have used the approximate equation, $\delta(E^{\gamma'}_{\bm{k}'}-E^{\gamma}_{\bm{k}}
\pm\hbar\omega_{\pm(\bm{k}-\bm{k}'),q_{z}})\simeq\delta(E^{\gamma'}_{\bm{k}'}-E^{\gamma}_{\bm{k}})$.
The iterative solution for Eq.~(\ref{eq:fmagbolintmain}) can be calculated, resulting in the following solution:
\begin{align}
f_{\rm C}(\bm{k},\gamma) &=\mathcal{G}_{\rm C}
(\bm{k},\gamma,\theta)
+
\frac{\hbar^{2}}{2m^{*}}
\int_{0}^{2\pi}\frac{d\varphi''}{2\pi}
\sum_{\gamma''} \gamma\gamma'' \int_{0}^{\infty}dk'' \, |\bm{k}''| \,
\nonumber \\
&\hspace{10mm}\times
\hat{\bm{h}}_{\rm eff}^T(\varphi)
\biggl{(}\hat{I}-\int_{0}^{2\pi} \! \! \frac{d\varphi'}{2\pi}\hat{\bm{h}}_{\rm eff}(\varphi')\cdot \hat{\bm{h}}_{\rm eff}^{T}(\varphi')\biggl{)}^{-1}
\hat{\bm{h}}_{\rm eff}(\varphi'') \delta(E^{\gamma''}_{\bm{k}''}-E^{\gamma}_{\bm{k}}) \mathcal{G}_{\rm C}(\bm{k}'',\gamma'',\theta). \label{eq:appGfmagC}
\end{align}
The modulation of the spin and current densities in the 2DEG is written as
\begin{align}
\Delta {\bm s}_{{\rm C}}(\theta)
&=\frac{\hbar}{4\pi}\sum_{\gamma}\int_{0}^{\infty}\!dk\,|\bm{k}|\int_{0}^{2\pi}\!\frac{d\varphi}{2\pi}
\langle \bm{k}\gamma|
\hat{\bm{\sigma}}
|\bm{k}\gamma\rangle f_{\rm C}(\bm{k},\gamma)
\label{def:sphiC}, \\
\Delta\bm{j}_{\rm C}(\theta)
&=\frac{e}{2\pi}\sum_{\gamma}
\int_{0}^{\infty}\!dk\,|\bm{k}|\int_{0}^{2\pi}\!\frac{d\varphi}{2\pi} \bm{v}(\bm{k},\gamma) f_{\rm C}(\bm{k},\gamma).\label{jREMR1C}
\end{align}
By substituting the solution of the distribution function given in Eq.~(\ref{eq:appGfmagC}) and using Eq.~(\ref{eq:pfpEdel}), the result of the main text given in Eqs.~(\ref{eq:JbJC})-(\ref{eq:bmgdef}) can be derived.
By using Eqs.~(\ref{eq:appGfmagC}) and (\ref{eq:fclean2}), we obtain the following analytical expression of $\delta \mu_{\rm C}(\varphi,\gamma)\equiv \delta\mu_{\rm C}(\bm{k},\gamma)|_{E^{\gamma}_{\bm{k}}=\mu}$ on the Fermi surface:
\begin{align}
\delta\mu_{\rm C}(\varphi,\gamma)=-\frac{4\pi D(\epsilon_{\rm F})S_0|\bar{T}|^2 \mathcal{A}eE_x \gamma}{\Gamma^2}\mathcal{J}(\varphi),\label{eq:delmuCJ}
\end{align}
where $\mathcal{J}(\varphi)$ is defined in Eq.~(\ref{eq:defmathJ}). Note that terms independent of $\gamma$ and $\theta$ are omitted in Eq.~(\ref{eq:delmuCJ}). 

To clarify the modulation of the distribution function, the chemical potential shift $\delta\mu_{\rm C}(\varphi,\gamma=+)$ is plotted in Fig.~\ref{fig:delmupC}, where we used the following constant for normalization:
\begin{align}
\mu_{x,{\rm C}}=  -\frac{4k_{\rm F} LD(\epsilon_{\rm F})S_{0}eE_{x}\mathcal{A}|\bar{T}|^{2} x}{\Gamma^{2}}, \quad (x=\alpha,\beta).
\end{align}
These plots are consistent with the schematic diagrams for the modulation of the distribution functions shown in Fig.~\ref{fig:Rashba}(f) and Fig.~\ref{fig:aob11}(f).
Note that, in the plot of Fig.~\ref{fig:delmupC} for $\alpha/\beta = 1.1$, the value at $\varphi = 3\pi/2$ is slightly larger than that at $\varphi = \pi$, and $|\Delta j_{{\rm C}y}|$ is slightly larger than $|\Delta j_{{\rm C}x}|$. This is thought to be because, at higher temperatures $T$, the scattering angle in magnon scattering induced by interfacial interactions tends to take larger values. On the other hand, at lower temperatures, only smaller scattering angles are possible, resulting in $|\Delta j_{{\rm C}x}|$ being larger than $|\Delta j_{{\rm C}y}|$.

\begin{figure*}[tb]
\centering
\includegraphics[width=145mm]{Fig8.pdf}
\caption{Left panels: The chemical potential shift
$\delta\mu_{\rm C}(\varphi,+)-\delta\mu_{\rm C}(\varphi_0,+)$ for the clean interface is plotted as a function of the azimuth angle $\varphi$ of the electron wavenumber for $\beta = 0$ (top plot) and $\alpha/\beta = 1.1$ (bottom plot).
The reference point of the chemical potential shift is taken as $\varphi_0=0,\pi$ for the former case, while as $\varphi_0=\pi/4,5\pi/4$ for the latter case.
Note that $\gamma=+$ corresponds to the inner Fermi surfaces.
The parameters are set as $k_{\rm B}T/\hbar\omega_{\bm 0}=3$, $|\gamma_{{\rm g}}|h_{\rm dc}/\omega_{\bm 0}=0.1$, and $k_{\rm F}a=0.1$. Right panels: Schematic diagrams for modulation of the distribution functions for $\beta = 0$ (the upper diagrams) and $\alpha/\beta = 1.1$ (the lower diagrams).
The orange (blue) regions represent the place where the distribution function increases (decreases) compared with the reference point.}
\label{fig:delmupC}
\end{figure*}

\end{widetext}

\bibliography{ref}

\end{document}